\documentclass[12pt]{article}

\usepackage{newtxtext,newtxmath}
\RequirePackage[colorlinks=true, allcolors=blue]{hyperref}

\usepackage{graphicx}
\usepackage{threeparttable}
\usepackage{booktabs}

\usepackage[letterpaper,margin=1in]{geometry}

\renewenvironment{abstract}
	{\quotation}
	{\endquotation}

\date{}

\makeatletter
\renewcommand{\fnum@figure}{\textbf{Figure \thefigure}}
\renewcommand{\fnum@table}{\textbf{Table \thetable}}

\newcommand{\kms}{km\,s$^{-1}$}
\newcommand{\arcsec}{$^{\prime\prime}$}
\newcommand{\arcmin}{$^{\prime}$}

\makeatother

\usepackage{scicite}

\usepackage{url}

\def\scititle{
	First Submillimeter Lights from Dome A: Tracing the Carbon Cycle in the Feedback of Massive Stars
}
\title{\bfseries \boldmath \scititle}

\author{
Yan Gong$^{1\dagger}$,
Jiaqiang Zhong$^{1,3\dagger}$,
Yuan Ren$^{1,3}$, 
Yilong Zhang$^{1,3}$, 
Daizhong Liu$^{1}$, \and
Yiping Ao$^{1,3}$, 
Qijun Yao$^{1}$, 
Wen Zhang$^{1,3}$, 
Wei Miao$^{1,3}$, 
Zhenhui Lin$^{1,3}$,  \and
Wenying Duan$^{1,2}$, 
Dong Liu$^{1}$, 
Kangmin Zhou$^{1}$, 
Jie Liu$^{1}$, 
Zheng Wang$^{1}$, 
Junda Jin$^{1}$, \and
Kun Zhang$^{1}$, 
Feng Wu$^{1}$, 
Jinpeng Li$^{1}$, 
Boliang Liu$^{1}$, 
Xuan Zhang$^{1,3}$, 
Zhengheng Luo$^{1,2}$, \and
Jiameng Wang$^{1,2}$, 
Huiqian Hao$^{4}$, 
Xingming Lu$^{4}$, 
Shaoming Xie$^{4}$, 
Jia Quan$^{5}$, \and
Yanjie Liu$^{5}$, 
Jingtao Liang$^{5}$, 
Xianjin Deng$^{6,7}$, 
Jun Jiang$^{6,7}$, 
Li Li$^{6,7}$, \and
Liang Guo$^{8}$, 
Tuo Ji$^{9}$, 
Peng Jiang$^{9}$, 
Yi Zhang$^{10}$, 
Chenggang Shu$^{10}$, \and
Sudeep Neupane$^{11}$, 
Ruiqing Mao$^{1}$, 
Shengcai Shi$^{1}$, 
Jing Li$^{1\ast}$ \and
	\small$^{1}$ Purple Mountain Observatory and Key Laboratory of Radio Astronomy, Chinese Academy of Sciences, 
    \and 
    \small Nanjing 210008, China 
    \and
	\small$^{2}$ School of Astronomy and Space Sciences, University of Science and Technology of China, Hefei 230026, China \and
    \small$^{3}$ State Key Laboratory of Radio Astronomy and Technology, National Astronomical Observatories, \and \small Chinese Academy of Sciences, Beijing 100101, China \and 
    \small$^{4}$ The 54th Research Institute of China Electronics Technology Group Corporation, Shijiazhuang 050081, China \and 
    \small$^{5}$ Technical Institute of Physics and Chemistry, Chinese Academy of Sciences, Beijing 100190, China \and 
    \small$^{6}$ Microsystem and Terahertz Research Center, China Academy of Engineering Physics, \and \small Chengdu, Sichuan 610200, China \and 
    \small$^{7}$ Institute of Electronic Engineering, China Academy of Engineering Physics, Mianyang, Sichuan 621999, China \and 
    \small$^{8}$ Changchun Institute of Optics, Fine Mechanics and Physics, Chinese Academy of Sciences, Changchun 130033, China \and 
    \small$^{9}$ Polar Research Institute of China; Key Laboratory for Polar Science, MNR, Shanghai 200136, China \and 
    \small$^{10}$ Shanghai Key Lab for Astrophysics, Shanghai Normal University, 100 Guilin Road, Shanghai 200234, China \and 
    \small$^{11}$ Max-Planck-Institut f{\"u}r Radioastronomie, Auf dem H{\"u}gel 69, D-53121 Bonn, Germany \and 
	\small$^\ast$Corresponding author. \quad Email: lijing@pmo.ac.cn    \and
	\small$^\dagger$These authors contributed equally to this work.
}

\begin{document} 

\maketitle

\begin{center}
\textbf{Teaser: First submillimeter maps from Antarctica’s Dome A reveal how massive stars power the cosmic carbon cycle.}
\end{center}
\vspace{1em}

\begin{abstract} \bfseries \boldmath
The cycling of carbon between its ionized, atomic, and molecular phases shapes the chemical compositions and physical conditions of the interstellar medium (ISM). However, ground-based studies of the full carbon cycle have been limited by atmospheric absorption. Dome~A, the most promising site for submillimeter astronomy, has long resisted successful submillimeter astronomical observations. Using the 60~cm Antarctic Terahertz Explorer, we present the first successful CO ($4-3$) and [CI] ($^3P_1 - ^3P_0$) mapping observations of two archetypal triggered massive star-formation regions at Dome~A. These data, together with archival [CII], provide the first complete characterization of all three carbon phases in these environments. We find elevated C$^{0}$/CO abundance ratios in high-extinction regions, plausibly driven by deep penetration of intense radiation fields from massive stars into a clumpy ISM. These findings mark a major milestone for submillimeter astronomy at Dome~A and offer valuable insights into the impact of massive star feedback on the surrounding ISM.

\end{abstract}

\noindent

\section*{Introduction}
Carbon, the second most abundant metal in the universe after oxygen\cite{2003ApJ...591.1220L}, plays a fundamental role in interstellar chemistry and the emergence of life on Earth\cite{1998Sci...282.2204H}. In the interstellar medium (ISM), carbon transitions between three primary phases: ionized (C$^{+}$), atomic (C$^{0}$), and molecular, predominantly in the form of carbon monoxide (CO). This phase cycling governs the thermal balance, molecular complexity, and star formation processes that drive the chemical evolution of galaxies. Known as the stellar nurseries in the ISM, molecular clouds are typically traced using line emission from low $J$ CO rotational transitions\cite{2001ApJ...547..792D,2013ARA&A..51..207B,2019ApJS..240....9S}, yet a substantial fraction of molecular gas extends beyond CO-bright regions\cite{2005Sci...307.1292G,2011A&A...536A..19P}. In the outer layers of molecular clouds, carbon predominantly exists as C$^{0}$ or C$^+$, while molecular hydrogen (H$_2$) is shielded from UV photodissociation by dust or self-shielding\cite{2010ApJ...716.1191W}. However, CO, more vulnerable to photodissociation, becomes depleted in UV-irradiated regions\cite{1988ApJ...334..771V}, leading to an H$_{2}$ component that is ``dark" in CO emission. With an ionization potential of 11.3~eV lower than hydrogen’s ionization potential of 13.6~eV, carbon is readily ionized by UV radiation, allowing C$^{+}$ to exist in both ionized and neutral gas. C$^{0}$, located between C$^{+}$ and CO in the carbon cycle, is abundant in the intermediate layers of PDRs where UV photons dissociate CO but are insufficient to ionize all carbon atoms. C$^{+}$ and C$^{0}$ thus offer complementary diagnostics to CO, tracing diffuse and CO-dark molecular gas. Therefore, a comprehensive view of carbon across its ionized, atomic, and molecular phases provides a powerful diagnostic of the composition, structure, and evolution of the ISM.

In classic PDR models\cite{2010ApJ...716.1191W,1999RvMP...71..173H,2022ARA&A..60..247W}, CO molecules are photodissociated into atomic carbon (C$^{0}$), which can be further ionized to C$^{+}$ under strong UV radiation. This suggests that C$^{0}$ and C$^{+}$ serve as unique tracers of environments inaccessible to CO. The [CII] 158~$\mu$m line, a key tracer of C$^{+}$, has been shown to reveal cloud kinematics that remain undetected in CO observations \cite{2019Natur.565..618P, 2021SciA....7.9511L, 2023NatAs...7..546S}, reinforcing this scenario. However, [CII] 158~$\mu$m emission can arise from molecular, atomic, and ionized gas\cite{1994ApJ...436..720H,2013A&A...554A.103P,2016A&A...591A..33R}, complicating its interpretation. In contrast, the two fine-structure transitions of C$^{0}$ in its ground state, $^{3}P_{1}-^3P_{0}$ (492~GHz) and $^{3}P_{2}-^3P_{1}$ (809~GHz), are powerful tracers of molecular gas\cite{2004MNRAS.351..147P,2014MNRAS.440L..81O,2015MNRAS.448.1607G,2022A&A...664A..80L}, with negligible contributions from atomic or ionized gas. This characteristic highlights the advantage of C$^{0}$ transitions in studying PDRs, which are crucial for understanding massive star feedback and cloud formation processes. However, observing [CI] emission remains challenging due to poor atmospheric transmission at these frequencies. While ground-based telescopes have successfully detected [CI] emissions, such observations remain limited compared to the extensive CO surveys. A submillimeter telescope at an exceptional observing site is therefore essential to enable a more comprehensive exploration of the carbon cycle on large scales in the ISM. This ambition has already driven the development of next-generation submillimeter observatories\cite{2018SPIE10700E..1MS,2020arXiv200807453S}.

\section*{Results and discussion}
\subsection*{The Road to First Submillimeter Lights at Dome A}
In 2005, the Chinese expedition successfully reached Dome A\cite{2010A&ARv..18..417B}, the summit of the East Antarctic ice sheet, a site considered highly promising for submillimeter observations\cite{2010PASP..122..490Y}. In 2009, the Chinese Kunlun Station was successfully established at Dome~A, providing a permanent research base for year-round scientific operations. Based on a remotely operated Fourier transform spectrometer deployed to Dome A during the 26th Chinese National Antarctica and Arctic Research Expedition (CHINARE), atmospheric radiation measurements across the full water vapor pure rotation band (0.75~THz to 15~THz) reveal substantial transmission within numerous frequency windows, underscoring the emergence of terahertz windows at this high-altitude Antarctic site \cite{2016NatAs...1E...1S}. Satellite data further confirm that Dome A exhibits the lowest median precipitable water vapor with the smallest fluctuation \cite{2017ApJ...848...64K}. Its excellent weather conditions fueled proposals for large-aperture terahertz telescopes and interferometers at the site \cite{2019arXiv190206398M}. However, astronomical observations at Dome~A pose significant challenges due to its remote location and harsh environmental conditions, including severe cold, persistent snow cover, and the absence of infrastructure. During the 24th CHINARE, the Pre-HEAT instrument, which is a 20 cm aperture submillimeter-wave telescope equipped with a 660~GHz Schottky diode heterodyne receiver \cite{2008SPIE.7012E..49K}, was employed to observe $^{13}$CO ($6-5$) for only one season at Dome~A\cite{2020RAA....20..168S}, but the observational results seem to be not available in the literature, leaving the conclusions elusive. While successful submillimeter observations have been made at other Antarctic locations, submillimeter observations at Dome A have yet to be fully realized.

\begin{figure}[!htbp]
\includegraphics[width=0.95\linewidth]{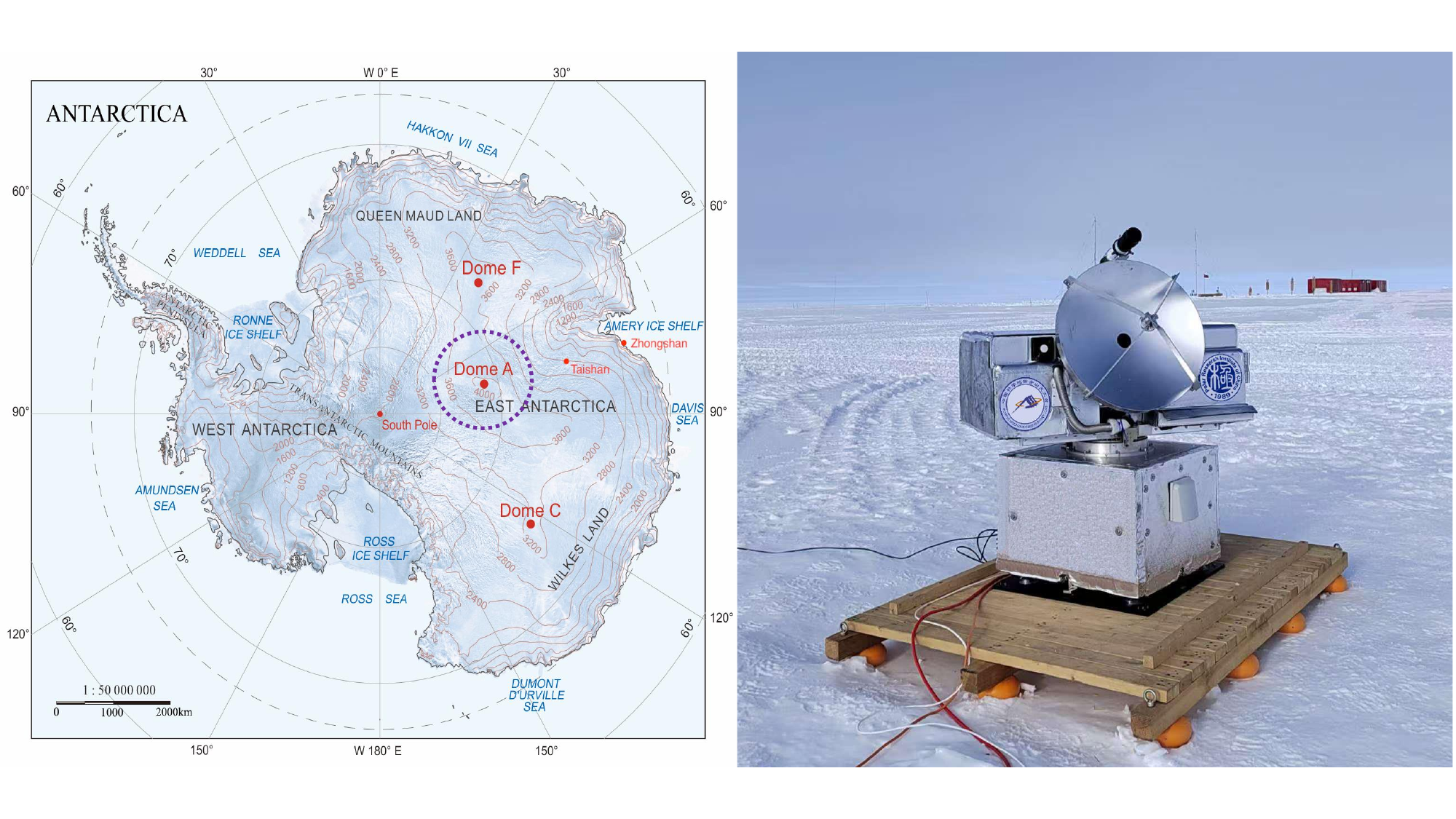}
\caption{Map of Antarctica with elevation contour lines\cite{2020RAA....20..168S}, with Dome A higlighted. Reproduced by permission of RAA. All rights reserved. The right inset shows a photo of ATE60 deployed at Dome A in January 2025 during the 41st CHINARE. The red building discernible in the distant background is the Chinese Kunlun Station.}
\label{fig:ate60}
\end{figure}

Due to its challenging environmental conditions, no submillimeter telescopes have been operational at Dome~A since the decommissioning of Pre-HEAT in 2008. However, thanks to the efforts of the 39th, 40th, and 41st expeditions of CHINARE, we had the opportunity to deploy portable submillimeter telescopes to this site. In particular, the Antarctic Terahertz Explorer with a 60~cm aperture (ATE60), a small-size submillimeter telescope equipped with a 460~GHz Nb-based Superconductor-Insulator-Superconductor (SIS) heterodyne receiver specifically designed for operation in polar conditions, was installed and operated at Dome~A in January 2025 during the 41st CHINARE (see Figure~\ref{fig:ate60}). SIS heterodyne receivers achieve substantially lower receiver temperatures than Schottky diode receivers \cite{research.0586} previously used for Pre-HEAT, providing a markedly improved opportunity for successful submillimeter observations under the extreme conditions at Dome~A. Despite the extremely harsh environment, we successfully carried out commissioning ATE60 observations of the CO ($4-3$) and [CI] ($^{3}P_{1}-^{3}P_{0}$) lines toward two archetypal triggered massive star-formation regions (i.e., RCW~79 and RCW~120) which exhibit prominent, well-define ring-like morphorlogies in their PDRs (see Figure~\ref{fig:3col}; more details are given in Section~\textit{Source Selection} of the supplementary materials). 
These observations represent a significant milestone for submillimeter astronomy at Dome~A, demonstrating the feasibility of high-frequency observations in one of the most challenging environments on Earth.

\subsection*{The Carbon Cycle and Stellar Feedback Revealed by ATE60}
The successful observations provide valuable CO ($4-3$) and [CI] ($^3P_1$–$^3P_0$) data for investigating physical conditions of the ISM. Figure~\ref{fig:cycle} presents the CO ($4-3$) and [CI] ($^{3}P_{1}-^{3}P_{0}$) results toward RCW~79 and RCW~120. Despite the existence of low-$J$ CO and [CII] 158~$\mu$m observations\cite{2023A&A...679L...5B,2015ApJ...806....7T,2021SciA....7.9511L,2017A&A...600A..93F,2018PASJ...70S..45O,2022A&A...659A..36K}, we present the first CO ($4-3$) and [CI] ($^{3}P_{1}-^{3}P_{0}$) mapping of RCW~79 and RCW~120 (see Section~\textit{Materials and Methods} in the supplementary materials for the details). In particular, the atomic carbon phase can be only investigated with [CI] data, while the CO ($4-3$) transition, with its higher upper-state energy and critical density compared to low-$J$ CO lines, offers enhanced sensitivity to warmer and denser gas components. Hence, our ATE60 observations of [CI] emission bridge a crucial gap in the carbon cycle, providing a complete picture of carbon phases in these regions. 

In both regions, bright [CI] and CO ($4$–$3$) emission is observed at the peripheries of the H{\scriptsize II} regions, peaking toward sites of second-generation star formation\cite{2006A&A...446..171Z, 2007A&A...472..835Z, 2017A&A...602A..95L}. In RCW~79, the peak positions of the [CII], [CI], and CO ($4-3$) emissions are spatially coincident (Figure~\ref{fig:cycle}A--\ref{fig:cycle}B), indicating a close association of these tracers in this region. In contrast, in RCW~120, both CO and [CI] emissions are confined to the southeastern part of the more extended [CII] distribution (Figure~\ref{fig:cycle}C--\ref{fig:cycle}D). These contrasting morphologies highlight the ability of [CII] to trace more diffuse and extended gas than either [CI] or CO. Moreover, [CI] emission shows a spatial distribution that closely follows that of CO, indicating that both trace similar molecular gas components. In particular, the [CI] emission exhibits a closer morphological correspondence with $^{13}$CO than with $^{12}$CO (Figure~\ref{fig:spectral}A--\ref{fig:spectral}B), which is likely attributed to the lower optical depths of the $^{13}$CO transitions relative to the $^{12}$CO transitions. The agreement between the [CI] and CO distributions is consistent with previous observations of the Orion A molecular cloud\cite{1999ApJ...527L..59I, 2013ApJ...774L..20S}, indicating that C$^{0}$ is not confined to the cloud surface but also traces the inner part of molecular clouds.

\begin{figure}[!htbp]
\includegraphics[width=0.95\linewidth]{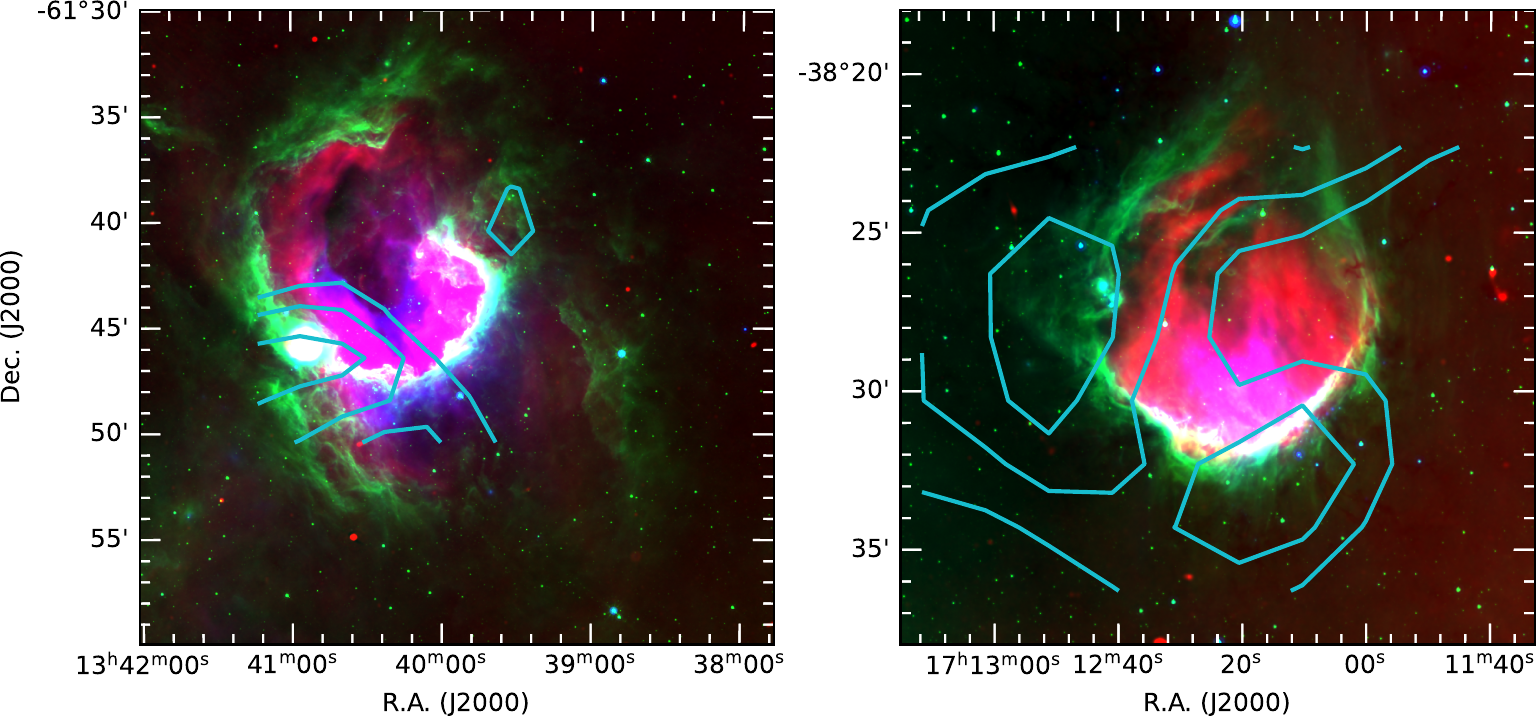}
\caption{\textbf{Three-color composite images of RCW~79 (left) and RCW~120 (right) overlaid with ATE60 [CI] integrated intensity contours.} The SARAO (South African Radio Astronomy Observatory) MeerKAT Galactic Plane Survey (SMGPS) 1.3~GHz radio continuum emission, Galactic Legacy Infrared Mid-Plane Survey Extraordinaire (GLIMPSE) 8~$\mu$m emission, and Multiband Imaging Photometer
for Spitzer Galactic Plane Survey (MIPSGAL) 24~$\mu$m emission are shown in red, green, and blue, respectively. For RCW~79 and RCW~120, the [CI] integrated intensity maps span velocity ranges of $-48$~\kms\,to $-43$~\kms\,and $-13$~\kms\,to $-3$~\kms, respectively. Contours are drawn at 4.5~K~km~s$^{-1}$ with increments of 1.5~K~kms$^{-1}$ for RCW~79, and at 15~K~km~s$^{-1}$ with increments of 4.5~K~kms$^{-1}$ for RCW~120.}
\label{fig:3col}
\end{figure}

\begin{figure}[!htbp]
\centering
\includegraphics[width=1.0\linewidth]{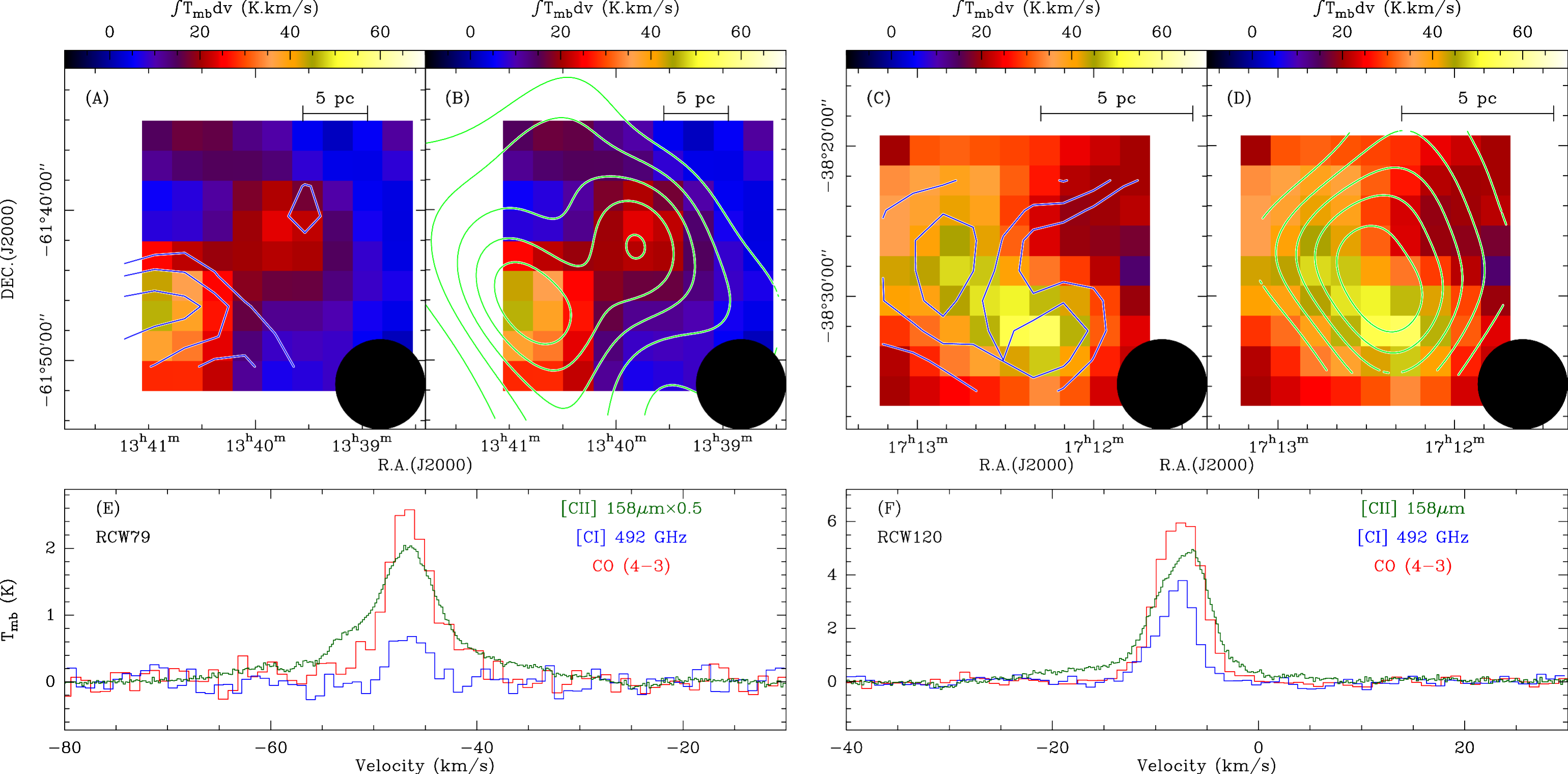}
\caption{\textbf{Distribution and spectra of C$^{+}$, C$^{0}$, and CO in RCW~79 and RCW~120.} (A) CO ($4-3$) integrated intensity map of RCW~79 overlaid with its [CI] integrated intensity contours (blue). The integrated velocity range for CO~($4-3$) is from $-40$~\kms\,to $-30$~\kms. The [CI] contours are the same as those in Figure~\ref{fig:3col}. (B) Similar to panel (A) but overlaid with the [CII] integrated intensity contours (green). The integrated velocity range for [CII] is $-70$~\kms\,to $-10$~\kms. The contours start at 25~K~\kms\,and increase in steps of 10~K~\kms. (C) and (D) similar to (A) and (B), respectively, but for RCW~120. The CO~($4$–$3$) and [CII] integrated velocity ranges are $-13$ to $-3$~km~s$^{-1}$ and $-30$ to $10$~km~s$^{-1}$, respectively; [CII] contours start at 25~K~km~s$^{-1}$ and increase by 10~K~km~s$^{-1}$. In panels (A)–(D), the color scale shows the CO ($4$–$3$) integrated intensity, and the beam size is indicated in the lower right corner. (E) [CII], [CI], and CO ($4-3$) spectra of RCW~79 averaged over the region indicated by the CO ($4-3$) integrated intensity map in panels (A)--(B). (F) Similar to panel (E) but for RCW~120. In panel (E), the [CII] spectrum has been scaled down for a better comparison, with the scaling factor provided in the legend. The [CII] 158~$\mu$m data are taken from the SOFIA legacy program FEEDBACK \cite{2021SciA....7.9511L,2023A&A...679L...5B,2020PASP..132j4301S}.}
\label{fig:cycle}
\end{figure}

Figures~\ref{fig:cycle}E and \ref{fig:cycle}F present the average spectra of [CII], [CI], and CO ($4-3$) toward RCW~79 and RCW~120. A prominent broad line wing is clearly visible in the [CII] spectra, whereas such features are absent in both [CI] and CO ($4-3$). This distinction is also evident in the spectra of the selected targets (see Figure~\ref{fig:spectral}C--\ref{fig:spectral}F). This indicates that [CII] emission can trace a distinct velocity component insensitive to both [CI] and CO, confirming that [CII] is an excellent probe of cloud dynamics \cite{2021SciA....7.9511L}. In RCW~120, a blue-shifted component at $\lesssim -$10~km~s$^{-1}$ is detected in [CI] and [CII] (see Figure~\ref{fig:cycle}F) and in CO, $^{13}$CO, [C I], and [C II] (Figure~\ref{fig:spectral}E), which likely originates from the near (front) side of the expanding shell \cite{2021SciA....7.9511L}. We find that the [CII]/[CI] and [CII]/CO line ratios in RCW~79 are about twice those measured in RCW~120. [CI] plays a crucial role in constraining the PDR models, as it bridges the transition between ionized and molecular gas layers and thus serves as a key diagnostic of the incident ultraviolet (UV) radiation field (see Figure~\ref{fig:pdr} in the supplementary materials). These line ratios are essential for determining the physical conditions within the PDRs, including the strength of the UV radiation field and the gas density. Using the combined diagnostics of the three carbon phases, we derive radiation fields of $G_{0}\sim170-250$~Habing for the two regions in RCW~79 and $G_{0}\sim 30$~Habing for the two regions in RCW~120 (see Table~\ref{tab:phy-info} and Section~\textit{PDR Models} in the supplementary materials). The modeling results are consistent with the interpretation that the elevated UV field in RCW~79 is produced by a cluster of multiple early-type O stars\cite{2020PASP..132j4301S}, whereas RCW~120, ionized by a single O8 V star\cite{2020PASP..132j4301S}, exhibits a comparatively weaker UV environment. This difference in the ionizing radiation field provides a straightforward explanation for the observed variations in the line ratios between the two regions.

Using observations of CO, $^{13}$CO, and C$^{0}$ transitions, we estimate the physical properties of the selected regions in RCW~79 and RCW~120 with a non-local thermodynamic equilibrium (non-LTE) analysis. The non-LTE modeling indicates gas temperatures of 14.8--21.5~K, H$_{2}$ number densities of $>10^{3}$~cm$^{-3}$, $^{13}$CO column densities of $\gtrsim 10^{16}$~cm$^{-2}$, and C$^{0}$-to-CO abundance ratios of $\gtrsim$0.29 (see Section~\textit{Non-LTE Analysis} in the supplementary materials). Adopting a [$^{12}$C/$^{13}$C] isotope ratio of 50\cite{2023A&A...670A..98Y}, a typical [CO/H$_{2}$] abundance of $8\times 10^{-5}$ \cite{1982ApJ...262..590F}, and the relation between the visual extinction and H$_{2}$ column densities\cite{1978ApJ...224..132B}, these cold regions are expected to have high extinction of $A_{\rm V} \gtrsim 7$. Therefore, our observations reveal C$^{0}$/CO abundance ratios of $\gtrsim$0.3 in cold and high-extinction regions ($A_{\rm V} \gtrsim 7$), which is independently confirmed by analysis using Hi-GAL–based H$_{2}$ column density maps (see Section~\textit{Extinction Map} in the supplementary materials). The derived C$^{0}$/CO abundance ratios appear to substantially exceed the typical values of $\lesssim$0.2 found both in the Milky Way disk at similar extinctions\cite{1999ApJ...527L..59I,2006ApJ...649..268S}. In nearby spiral galaxies, C$^{0}$/CO abundance ratios are typically $\sim$0.1 across most of the disks of NGC~3627 and NGC~4321, with enhancements up to $\sim$1 in in NGC~1808's starburst and even exceeding $1-5$ in NGC~7469's strong active galactic nucleus environments\cite{2023A&A...672A..36L}. Our measured values are therefore significantly higher than the representative ``normal-disk" level commonly found in the Milky Way and nearby spiral galaxies, suggesting an enhanced reservoir of atomic carbon in these regions.

Three primary scenarios may account for the elevated C$^{0}$/CO abundance ratios. One possibility is that high cosmic ray ionization rates can enhance C$^{0}$ abundances\cite{2015ApJ...803...37B}, but the lack of prominent cosmic ray sources in these environments makes this explanation unlikely. Alternatively, in the early stages of molecular cloud formation, the conversion from atomic to molecular gas proceeds with CO forming more slowly than C$^{0}$, elevating the C$^{0}$/CO abundance ratio in young clouds\cite{2004ApJ...612..921B}. However, as the clouds examined here already host massive stars, this scenario is also disfavored. Instead, in evolved molecular clouds exposed to intense UV radiation, CO is efficiently photodissociated into C$^{0}$ by UV photons\cite{1999RvMP...71..173H}. The elevated C$^{0}$/CO abundance ratios in our targets are thus more plausibly explained by UV-driven CO dissociation resulting from strong UV radiation fields of nearby massive stars. Comparison with PDR models (see Section~\textit{PDR Models} in the supplementary materials) further indicates that a clumpy PDR structure is required, as it enables deep penetration of UV photons necessary to reproduce the observed abundance ratios. Therefore, the pronounced elevation in the C$^{0}$/CO abundance ratio is likely driven by enhanced photodissociation within a clumpy ISM under the influence of stellar feedback, underscoring the important role of massive stars in shaping the chemical composition of the surrounding ISM.

\subsection*{Prospect}
Our CO ($4-3$) and [CI] ($^3P_1 - ^3P_0$) mapping observations of RCW~79 and RCW~120, conducted with ATE60, represent the first successful submillimeter detections from Dome~A. These measurements offer a complete view of the carbon cycle in environments shaped by massive star feedback. Furthermore, Dome~A is characterized by a typical winter precipitable water vapor of $<$0.2~mm, as indicated by early Fourier transform spectrometer and satellite measurements\cite{2016NatAs...1E...1S,2017ApJ...848...64K}. These results demonstrate the extraordinary scientific promise of Dome~A as a platform for submillimeter and terahertz astronomy. The findings not only underscore the profound influence of stellar feedback from massive stars on the chemical compositions in their surrounding ISM, but also establish a foundation for future large-scale, high-angular-resolution surveys of key terahertz tracers (e.g., [CI], [NII], H$_{2}$D$^{+}$, and high-$J$ CO transitions) which can probe the cold and dynamic Universe in unprecedented detail. Building on earlier pioneering efforts at other Antarctic sites, including those from the Antarctic Submillimeter Telescope and Remote Observatory (AST/RO\cite{1997ApJ...480L..59S,2001ApJ...548..253O,2001ApJ...553..274Z}) as well as the High Elevation Antarctic Terahertz telescope (HEAT\cite{2015ApJ...811...13B}), this milestone further affirms the unique scientific advantages of the Antarctic plateau for studying the interplay of chemistry, radiation, and star formation in the cosmos, highlighting the importance of advancing Antarctic astronomy to fully realize its exceptional potential for submillimeter and terahertz science.

\begin{figure}[!htbp]
\includegraphics[width=0.95\linewidth]{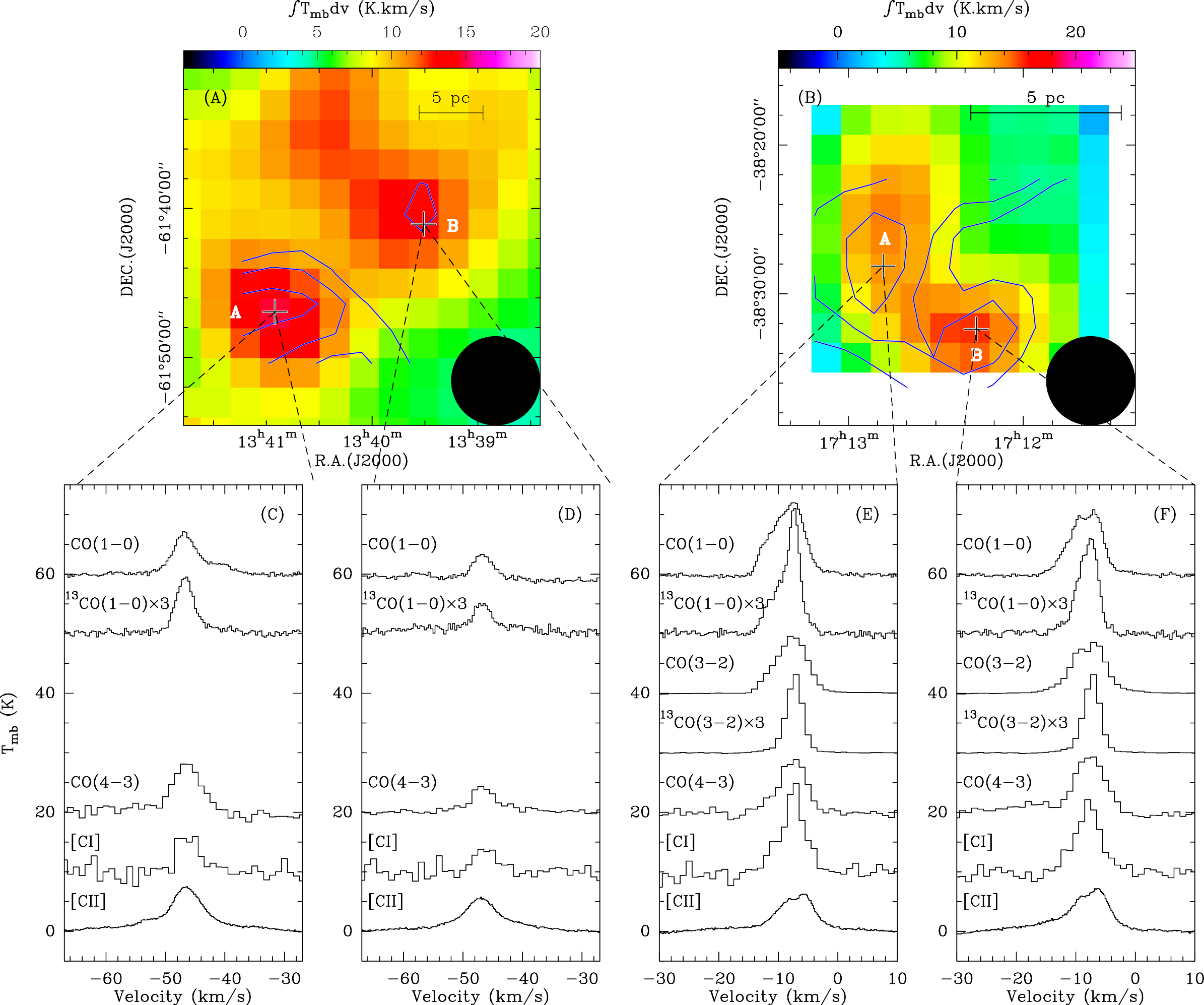}
\caption{\textbf{Observed spectra of the selected targets.} (A) Integrated-intensity map of $^{13}$CO ($1-0$) overlaid with [CI] contours for RCW~79. The $^{13}$CO ($1-0$) integrated velocity range is from $-$60~\kms\,to $-$30~\kms. (B) Integrated-intensity map of $^{13}$CO ($3-2$) overlaid with [CI] contours for RCW~120. The $^{13}$CO ($3-2$) integrated velocity range is from $-$13~\kms\,to $-$3~\kms. In panels (A) and (B), the beam sizes are indicated in the lower right corner and the [CI] contours are the same as in Figure~\ref{fig:cycle}. Panels (C)--(F) present observed spectra of the selected targets which are extracted from the pixels indicated by pluses in panels (A) and (B). The transitions are labeled in each panel and the $^{13}$CO transitions are scaled by a factor of 3 for comparison.}
\label{fig:spectral}
\end{figure}

\section*{Materials and Methods}\label{sec:method}
To enable submillimeter observations in the challenging Antarctic environment, ATE60 was developed with low-power consumption and a miniaturized receiver design. ATE60 was deployed to Dome A during the 41st CHINARE in the Antarctic summer (see Figure~\ref{fig:ate60}). This telescope was positioned at a longitude of 77$^{\circ}$06$^{\prime}$38.82$^{\prime\prime}$ and a latitude of $-$81$^{\circ}$25$^{\prime}$01.63$^{\prime\prime}$, and an altitude of 4093~m. ATE60 is equipped with a single-pixel Nb-based Superconductor-Insulator-Superconductor (SIS) heterodyne receiver, operating in the double-sideband mode across the 456--504~GHz frequency range. Data processing was facilitated by fast Fourier transform spectrometers (FFTSs). The FFTSs, offering 32768 channels, cover an intermediate frequency bandwidth of 2.4~GHz. This configuration resulted in a spectral resolution of $\sim$73~kHz, corresponding to a velocity spacing of $\sim$0.05~\kms\,at 460~GHz. During operation, ATE60 requires only 1.5–2~kW of power, supplied by diesel generators housed in a dedicated power cabin that also supports the broader activities of the expedition team. Shortly after the successful relocation of ATE60, we first validated our receiver system by observing the massive star formation region NGC~6334I. The immediate detection of CO ($4-3$) and [CI] ($^{3}P_{1}-^{3}P_{0}$) at the expected velocities confirmed the observing capability of ATE60. Building on this success, we subsequently conducted CO ($4-3$) and [CI] ($^{3}P_{1}-^{3}P_{0}$) observations toward RCW~79 and RCW~120 between 6 and 17 January 2025, during the expedition at the Dome A site. 

Based on skydip measurements taken during the observations, the typical atmospheric optical depth at 460~GHz is $\sim$0.5, corresponding to a precipitable water vapor (PWV) of $\sim$1~mm. These conditions are comparable to the median PWV value ($\sim$1.2~mm) during the Bolivian winter at Llano de Chajnantor, the site of the Atacama Pathfinder EXperiment (APEX) and Atacama Large Millimeter/submillimeter Array\cite{2006A&A...454L..13G}. The two-load (hot and cold) calibration was used to set the antenna temperature scale. During the observations, ambient temperatures were about 240~K (i.e., $-$33$^{\circ}$C). The receiver temperatures were about 600~K and 300~K, with the corresponding atmosphere-corrected system temperatures estimated at 1093--3921K and 1120--3750~K at 462~GHz and 493~GHz, respectively. The absolute flux calibration uncertainties were assumed to be 20\% (see Section~\textit{Flux Calibration} in the supplementary materials). The half-power beam widths (HPBWs) were measured to be approximately 270\arcsec\,and 240\arcsec\,at 460~GHz and 492~GHz, respectively. The overall pointing error was estimated to be $\lesssim$1$^{\prime}$ (see Section~\textit{Pointing} in the supplementary materials). 

After initial mapping of RCW~79 and RCW~120, we observed that RCW~120 exhibits brighter [CI] ($^{3}P_{1}-^{3}P_{0}$) emission compared to RCW~79 (see Figure~\ref{fig:cycle}). Consequently, we prioritized our ATE60 CI observations on RCW~120 to improve sensitivity. To enhance the signal-to-noise ratio, we smoothed the spectra to a velocity spacing of 1~\kms, and the noise level of [CI] data at this channel width is $\sim$0.15~K for RCW~79 and $\sim$0.12~K for RCW~120 on the antenna temperature scale.


\clearpage 

%
\bibliography{ref} 

@ARTICLE{2003ApJ...591.1220L,
       author = {{Lodders}, Katharina},
        title = "{Solar System Abundances and Condensation Temperatures of the Elements}",
      journal = {\apj},
         year = 2003,
        month = jul,
       volume = {591},
       number = {2},
        pages = {1220-1247},
          doi = {10.1086/375492},
       adsurl = {https://ui.adsabs.harvard.edu/abs/2003ApJ...591.1220L}
}

@ARTICLE{1998Sci...282.2204H,
       author = {{Henning}, T. and {Salama}, F.},
        title = "{Carbon in the Universe}",
      journal = {Science},
         year = 1998,
        month = dec,
       volume = {282},
        pages = {2204},
          doi = {10.1126/science.282.5397.2204},
       adsurl = {https://ui.adsabs.harvard.edu/abs/1998Sci...282.2204H}
}

@ARTICLE{2024MNRAS.531..649G,
       author = {{Goedhart}, S. and {Cotton}, W.~D. and {Camilo}, F. and {Thompson}, M.~A. and {Umana}, G. and {Bietenholz}, M. and {Woudt}, P.~A. and {Anderson}, L.~D. and {Bordiu}, C. and {Buckley}, D.~A.~H. and {Buemi}, C.~S. and {Bufano}, F. and {Cavallaro}, F. and {Chen}, H. and {Chibueze}, J.~O. and {Egbo}, D. and {Frank}, B.~S. and {Hoare}, M.~G. and {Ingallinera}, A. and {Irabor}, T. and {Kraan-Korteweg}, R.~C. and {Kurapati}, S. and {Leto}, P. and {Loru}, S. and {Mutale}, M. and {Obonyo}, W.~O. and {Plavin}, A. and {Rajohnson}, S.~H.~A. and {Rigby}, A. and {Riggi}, S. and {Seidu}, M. and {Serra}, P. and {Smart}, B.~M. and {Stappers}, B.~W. and {Steyn}, N. and {Surnis}, M. and {Trigilio}, C. and {Williams}, G.~M. and {Abbott}, T.~D. and {Adam}, R.~M. and {Asad}, K.~M.~B. and {Baloyi}, T. and {Bauermeister}, E.~F. and {Bennet}, T.~G.~H. and {Bester}, H. and {Botha}, A.~G. and {Brederode}, L.~R.~S. and {Buchner}, S. and {Burger}, J.~P. and {Cheetham}, T. and {Cloete}, K. and {de Villiers}, M.~S. and {de Villiers}, D.~I.~L. and {du Toit}, L.~J. and {Esterhuyse}, S.~W.~P. and {Fanaroff}, B.~L. and {Fourie}, D.~J. and {Gamatham}, R.~R.~G. and {Gatsi}, T.~G. and {Geyer}, M. and {Gouws}, M. and {Gumede}, S.~C. and {Heywood}, I. and {Hokwana}, A. and {Hoosen}, S.~W. and {Horn}, D.~M. and {Horrell}, L.~M.~G. and {Hugo}, B.~V. and {Isaacson}, A.~I. and {J{\'o}zsa}, G.~I.~G. and {Jonas}, J.~L. and {Jordaan}, J.~D.~B.~L. and {Joubert}, A.~F. and {Julie}, R.~P.~M. and {Kapp}, F.~B. and {Kriek}, N. and {Kriel}, H. and {Krishnan}, V.~K. and {Kusel}, T.~W. and {Legodi}, L.~S. and {Lehmensiek}, R. and {Lord}, R.~T. and {Macfarlane}, P.~S. and {Magnus}, L.~G. and {Magozore}, C. and {Main}, J.~P.~L. and {Malan}, J.~A. and {Manley}, J.~R. and {Marais}, S.~J. and {Maree}, M.~D.~J. and {Martens}, A. and {Maruping}, P. and {McAlpine}, K. and {Merry}, B.~C. and {Mgodeli}, M. and {Millenaar}, R.~P. and {Mokone}, O.~J. and {Monama}, T.~E. and {New}, W.~S. and {Ngcebetsha}, B. and {Ngoasheng}, K.~J. and {Nicolson}, G.~D. and {Ockards}, M.~T. and {Oozeer}, N. and {Passmoor}, S.~S. and {Patel}, A.~A. and {Peens-Hough}, A. and {Perkins}, S.~J. and {Ramaila}, A.~J.~T. and {Ratcliffe}, S.~M. and {Renil}, R. and {Richter}, L.~L. and {Salie}, S. and {Sambu}, N. and {Schollar}, C.~T.~G. and {Schwardt}, L.~C. and {Schwartz}, R.~L. and {Serylak}, M. and {Siebrits}, R. and {Sirothia}, S.~K. and {Slabber}, M.~J. and {Smirnov}, O.~M. and {Tiplady}, A.~J. and {van Balla}, T.~J. and {van der Byl}, A. and {Van Tonder}, V. and {Venter}, A.~J. and {Venter}, M. and {Welz}, M.~G. and {Williams}, L.~P.},
        title = "{The SARAO MeerKAT 1.3 GHz Galactic Plane Survey}",
      journal = {\mnras},
         year = 2024,
        month = jun,
       volume = {531},
       number = {1},
        pages = {649-681},
          doi = {10.1093/mnras/stae1166},
archivePrefix = {arXiv},
       eprint = {2312.07275},
 primaryClass = {astro-ph.GA},
       adsurl = {https://ui.adsabs.harvard.edu/abs/2024MNRAS.531..649G}
}

@ARTICLE{2003PASP..115..953B,
       author = {{Benjamin}, Robert A. and {Churchwell}, E. and {Babler}, Brian L. and {Bania}, T.~M. and {Clemens}, Dan P. and {Cohen}, Martin and {Dickey}, John M. and {Indebetouw}, R{\'e}my and {Jackson}, James M. and {Kobulnicky}, Henry A. and {Lazarian}, Alex and {Marston}, A.~P. and {Mathis}, John S. and {Meade}, Marilyn R. and {Seager}, Sara and {Stolovy}, S.~R. and {Watson}, C. and {Whitney}, Barbara A. and {Wolff}, Michael J. and {Wolfire}, Mark G.},
        title = "{GLIMPSE. I. An SIRTF Legacy Project to Map the Inner Galaxy}",
      journal = {\pasp},
         year = 2003,
        month = aug,
       volume = {115},
       number = {810},
        pages = {953-964},
          doi = {10.1086/376696},
archivePrefix = {arXiv},
       eprint = {astro-ph/0306274},
 primaryClass = {astro-ph},
       adsurl = {https://ui.adsabs.harvard.edu/abs/2003PASP..115..953B}
}

@ARTICLE{2009PASP..121...76C,
       author = {{Carey}, S.~J. and {Noriega-Crespo}, A. and {Mizuno}, D.~R. and {Shenoy}, S. and {Paladini}, R. and {Kraemer}, K.~E. and {Price}, S.~D. and {Flagey}, N. and {Ryan}, E. and {Ingalls}, J.~G. and {Kuchar}, T.~A. and {Pinheiro Gon{\c{c}}alves}, Daniela and {Indebetouw}, R. and {Billot}, N. and {Marleau}, F.~R. and {Padgett}, D.~L. and {Rebull}, L.~M. and {Bressert}, E. and {Ali}, Babar and {Molinari}, S. and {Martin}, P.~G. and {Berriman}, G.~B. and {Boulanger}, F. and {Latter}, W.~B. and {Miville-Deschenes}, M.~A. and {Shipman}, R. and {Testi}, L.},
        title = "{MIPSGAL: A Survey of the Inner Galactic Plane at 24 and 70 {\ensuremath{\mu}}m}",
      journal = {\pasp},
         year = 2009,
        month = jan,
       volume = {121},
       number = {875},
        pages = {76},
          doi = {10.1086/596581},
       adsurl = {https://ui.adsabs.harvard.edu/abs/2009PASP..121...76C}
}

@ARTICLE{2015MNRAS.454.4282M,
       author = {{Marsh}, K.~A. and {Whitworth}, A.~P. and {Lomax}, O.},
        title = "{Temperature as a third dimension in column-density mapping of dusty astrophysical structures associated with star formation}",
      journal = {\mnras},
         year = 2015,
        month = dec,
       volume = {454},
       number = {4},
        pages = {4282-4292},
          doi = {10.1093/mnras/stv2248},
archivePrefix = {arXiv},
       eprint = {1509.08699},
 primaryClass = {astro-ph.IM},
       adsurl = {https://ui.adsabs.harvard.edu/abs/2015MNRAS.454.4282M}
}

@ARTICLE{1994ApJ...436..720H,
       author = {{Heiles}, Carl},
        title = "{On the Origin of the Diffuse C + 158 Micron Line Emission}",
      journal = {\apj},
         year = 1994,
        month = dec,
       volume = {436},
        pages = {720},
          doi = {10.1086/174945},
       adsurl = {https://ui.adsabs.harvard.edu/abs/1994ApJ...436..720H}
}

@ARTICLE{2016A&A...591A..33R,
       author = {{R{\"o}llig}, M. and {Simon}, R. and {G{\"u}sten}, R. and {Stutzki}, J. and {Israel}, F.~P. and {Jacobs}, K.},
        title = "{[C II] 158 {\ensuremath{\mu}}m and [N II] 205 {\ensuremath{\mu}}m emission from IC 342. Disentangling the emission from ionized and photo-dissociated regions}",
      journal = {\aap},
         year = 2016,
        month = jun,
       volume = {591},
          eid = {A33},
        pages = {A33},
          doi = {10.1051/0004-6361/201526267},
archivePrefix = {arXiv},
       eprint = {1604.07362},
 primaryClass = {astro-ph.GA},
       adsurl = {https://ui.adsabs.harvard.edu/abs/2016A&A...591A..33R}
}

@ARTICLE{2013A&A...554A.103P,
       author = {{Pineda}, J.~L. and {Langer}, W.~D. and {Velusamy}, T. and {Goldsmith}, P.~F.},
        title = "{A Herschel [C ii] Galactic plane survey. I. The global distribution of ISM gas components}",
      journal = {\aap},
         year = 2013,
        month = jun,
       volume = {554},
          eid = {A103},
        pages = {A103},
          doi = {10.1051/0004-6361/201321188},
archivePrefix = {arXiv},
       eprint = {1304.7770},
 primaryClass = {astro-ph.GA},
       adsurl = {https://ui.adsabs.harvard.edu/abs/2013A&A...554A.103P}
}

@ARTICLE{2014MNRAS.440L..81O,
       author = {{Offner}, S.~S.~R. and {Bisbas}, T.~G. and {Bell}, T.~A. and {Viti}, S.},
        title = "{An alternative accurate tracer of molecular clouds: the `XCi-factor'.}",
      journal = {\mnras},
         year = 2014,
        month = may,
       volume = {440},
        pages = {L81-L85},
          doi = {10.1093/mnrasl/slu013},
archivePrefix = {arXiv},
       eprint = {1401.5072},
 primaryClass = {astro-ph.SR},
       adsurl = {https://ui.adsabs.harvard.edu/abs/2014MNRAS.440L..81O}
}

@ARTICLE{2015MNRAS.448.1607G,
       author = {{Glover}, Simon C.~O. and {Clark}, Paul C. and {Micic}, Milica and {Molina}, Faviola},
        title = "{Modelling [C I] emission from turbulent molecular clouds}",
      journal = {\mnras},
         year = 2015,
        month = apr,
       volume = {448},
       number = {2},
        pages = {1607-1627},
          doi = {10.1093/mnras/stu2699},
archivePrefix = {arXiv},
       eprint = {1403.3530},
 primaryClass = {astro-ph.GA},
       adsurl = {https://ui.adsabs.harvard.edu/abs/2015MNRAS.448.1607G}
}

@ARTICLE{2022A&A...664A..80L,
       author = {{Lee}, M. -Y. and {Wyrowski}, F. and {Menten}, K. and {Tiwari}, M. and {G{\"u}sten}, R.},
        title = "{ATLASGAL-selected massive clumps in the inner Galaxy. X. Observations of atomic carbon at 492 GHz}",
      journal = {\aap},
         year = 2022,
        month = aug,
       volume = {664},
          eid = {A80},
        pages = {A80},
          doi = {10.1051/0004-6361/202142404},
archivePrefix = {arXiv},
       eprint = {2204.11414},
 primaryClass = {astro-ph.GA},
       adsurl = {https://ui.adsabs.harvard.edu/abs/2022A&A...664A..80L}
}

@ARTICLE{2019ApJ...885..131R,
       author = {{Reid}, M.~J. and {Menten}, K.~M. and {Brunthaler}, A. and {Zheng}, X.~W. and {Dame}, T.~M. and {Xu}, Y. and {Li}, J. and {Sakai}, N. and {Wu}, Y. and {Immer}, K. and {Zhang}, B. and {Sanna}, A. and {Moscadelli}, L. and {Rygl}, K.~L.~J. and {Bartkiewicz}, A. and {Hu}, B. and {Quiroga-Nu{\~n}ez}, L.~H. and {van Langevelde}, H.~J.},
        title = "{Trigonometric Parallaxes of High-mass Star-forming Regions: Our View of the Milky Way}",
      journal = {\apj},
         year = 2019,
        month = nov,
       volume = {885},
       number = {2},
          eid = {131},
        pages = {131},
          doi = {10.3847/1538-4357/ab4a11},
archivePrefix = {arXiv},
       eprint = {1910.03357},
 primaryClass = {astro-ph.GA},
       adsurl = {https://ui.adsabs.harvard.edu/abs/2019ApJ...885..131R}
}

@ARTICLE{2017MNRAS.471.2730M,
       author = {{Marsh}, K.~A. and {Whitworth}, A.~P. and {Lomax}, O. and {Ragan}, S.~E. and {Becciani}, U. and {Cambr{\'e}sy}, L. and {Di Giorgio}, A. and {Eden}, D. and {Elia}, D. and {Kacsuk}, P. and {Molinari}, S. and {Palmeirim}, P. and {Pezzuto}, S. and {Schneider}, N. and {Sciacca}, E. and {Vitello}, F.},
        title = "{Multitemperature mapping of dust structures throughout the Galactic Plane using the PPMAP tool with Herschel Hi-GAL data}",
      journal = {\mnras},
         year = 2017,
        month = nov,
       volume = {471},
       number = {3},
        pages = {2730-2742},
          doi = {10.1093/mnras/stx1723},
archivePrefix = {arXiv},
       eprint = {1707.03808},
 primaryClass = {astro-ph.GA},
       adsurl = {https://ui.adsabs.harvard.edu/abs/2017MNRAS.471.2730M}
}

@ARTICLE{2001ApJ...547..792D,
       author = {{Dame}, T.~M. and {Hartmann}, Dap and {Thaddeus}, P.},
        title = "{The Milky Way in Molecular Clouds: A New Complete CO Survey}",
      journal = {\apj},
         year = 2001,
        month = feb,
       volume = {547},
       number = {2},
        pages = {792-813},
          doi = {10.1086/318388},
archivePrefix = {arXiv},
       eprint = {astro-ph/0009217},
 primaryClass = {astro-ph},
       adsurl = {https://ui.adsabs.harvard.edu/abs/2001ApJ...547..792D}
}

@ARTICLE{2025ApJS..280...31B,
       author = {{Barnes}, Peter J. and {Barnes}, Dylan G.~H. and {Hern{\'a}ndez}, Audra K. and {Lopez}, Sebastian and {Muller}, Erik},
        title = "{The Three-mm Ultimate Mopra Milky Way Survey. III. Data Release 6, An Atlas of Physical Conditions, Global Mass Conversion Laws, and 3D Physical Architecture of the Molecular Interstellar Medium in the Fourth Quadrant}",
      journal = {\apjs},
         year = 2025,
        month = sep,
       volume = {280},
       number = {1},
          eid = {31},
        pages = {31},
          doi = {10.3847/1538-4365/adf32b},
archivePrefix = {arXiv},
       eprint = {2503.04887},
 primaryClass = {astro-ph.GA},
       adsurl = {https://ui.adsabs.harvard.edu/abs/2025ApJS..280...31B}
}

@ARTICLE{2013ARA&A..51..207B,
       author = {{Bolatto}, Alberto D. and {Wolfire}, Mark and {Leroy}, Adam K.},
        title = "{The CO-to-H$_{2}$ Conversion Factor}",
      journal = {\araa},
         year = 2013,
        month = aug,
       volume = {51},
       number = {1},
        pages = {207-268},
          doi = {10.1146/annurev-astro-082812-140944},
archivePrefix = {arXiv},
       eprint = {1301.3498},
 primaryClass = {astro-ph.GA},
       adsurl = {https://ui.adsabs.harvard.edu/abs/2013ARA&A..51..207B}
}

@ARTICLE{2019ApJS..240....9S,
       author = {{Su}, Yang and {Yang}, Ji and {Zhang}, Shaobo and {Gong}, Yan and {Wang}, Hongchi and {Zhou}, Xin and {Wang}, Min and {Chen}, Zhiwei and {Sun}, Yan and {Chen}, Xuepeng and {Xu}, Ye and {Jiang}, Zhibo},
        title = "{The Milky Way Imaging Scroll Painting (MWISP): Project Details and Initial Results from the Galactic Longitudes of 25.{\textdegree}8-49.{\textdegree}7}",
      journal = {\apjs},
         year = 2019,
        month = jan,
       volume = {240},
       number = {1},
          eid = {9},
        pages = {9},
          doi = {10.3847/1538-4365/aaf1c8},
archivePrefix = {arXiv},
       eprint = {1901.00285},
 primaryClass = {astro-ph.GA},
       adsurl = {https://ui.adsabs.harvard.edu/abs/2019ApJS..240....9S}
}

@ARTICLE{2004MNRAS.351..147P,
       author = {{Papadopoulos}, P.~P. and {Thi}, W. -F. and {Viti}, S.},
        title = "{CI lines as tracers of molecular gas, and their prospects at high redshifts}",
      journal = {\mnras},
         year = 2004,
        month = jun,
       volume = {351},
       number = {1},
        pages = {147-160},
          doi = {10.1111/j.1365-2966.2004.07762.x},
archivePrefix = {arXiv},
       eprint = {astro-ph/0403092},
 primaryClass = {astro-ph},
       adsurl = {https://ui.adsabs.harvard.edu/abs/2004MNRAS.351..147P}
}

@ARTICLE{2005Sci...307.1292G,
       author = {{Grenier}, Isabelle A. and {Casandjian}, Jean-Marc and {Terrier}, R{\'e}gis},
        title = "{Unveiling Extensive Clouds of Dark Gas in the Solar Neighborhood}",
      journal = {Science},
         year = 2005,
        month = feb,
       volume = {307},
       number = {5713},
        pages = {1292-1295},
          doi = {10.1126/science.1106924},
       adsurl = {https://ui.adsabs.harvard.edu/abs/2005Sci...307.1292G}
}

@ARTICLE{2011A&A...536A..19P,
       author = {{Planck Collaboration} and {Ade}, P.~A.~R. and {Aghanim}, N. and {Arnaud}, M. and {Ashdown}, M. and {Aumont}, J. and {Baccigalupi}, C. and {Balbi}, A. and {Banday}, A.~J. and {Barreiro}, R.~B. and {Bartlett}, J.~G. and {Battaner}, E. and {Benabed}, K. and {Beno{\^\i}t}, A. and {Bernard}, J. -P. and {Bersanelli}, M. and {Bhatia}, R. and {Bock}, J.~J. and {Bonaldi}, A. and {Bond}, J.~R. and {Borrill}, J. and {Bouchet}, F.~R. and {Boulanger}, F. and {Bucher}, M. and {Burigana}, C. and {Cabella}, P. and {Cardoso}, J. -F. and {Catalano}, A. and {Cay{\'o}n}, L. and {Challinor}, A. and {Chamballu}, A. and {Chiang}, L. -Y. and {Chiang}, C. and {Christensen}, P.~R. and {Clements}, D.~L. and {Colombi}, S. and {Couchot}, F. and {Coulais}, A. and {Crill}, B.~P. and {Cuttaia}, F. and {Dame}, T.~M. and {Danese}, L. and {Davies}, R.~D. and {Davis}, R.~J. and {de Bernardis}, P. and {de Gasperis}, G. and {de Rosa}, A. and {de Zotti}, G. and {Delabrouille}, J. and {Delouis}, J. -M. and {D{\'e}sert}, F. -X. and {Dickinson}, C. and {Dobashi}, K. and {Donzelli}, S. and {Dor{\'e}}, O. and {D{\"o}rl}, U. and {Douspis}, M. and {Dupac}, X. and {Efstathiou}, G. and {En{\ss}lin}, T.~A. and {Eriksen}, H.~K. and {Falgarone}, E. and {Finelli}, F. and {Forni}, O. and {Fosalba}, P. and {Frailis}, M. and {Franceschi}, E. and {Fukui}, Y. and {Galeotta}, S. and {Ganga}, K. and {Giard}, M. and {Giardino}, G. and {Giraud-H{\'e}raud}, Y. and {Gonz{\'a}lez-Nuevo}, J. and {G{\'o}rski}, K.~M. and {Gratton}, S. and {Gregorio}, A. and {Grenier}, I.~A. and {Gruppuso}, A. and {Hansen}, F.~K. and {Harrison}, D. and {Helou}, G. and {Henrot-Versill{\'e}}, S. and {Herranz}, D. and {Hildebrandt}, S.~R. and {Hivon}, E. and {Hobson}, M. and {Holmes}, W.~A. and {Hovest}, W. and {Hoyland}, R.~J. and {Huffenberger}, K.~M. and {Jaffe}, A.~H. and {Jones}, W.~C. and {Juvela}, M. and {Kawamura}, A. and {Keih{\"a}nen}, E. and {Keskitalo}, R. and {Kisner}, T.~S. and {Kneissl}, R. and {Knox}, L. and {Kurki-Suonio}, H. and {Lagache}, G. and {Lamarre}, J. -M. and {Lasenby}, A. and {Laureijs}, R.~J. and {Lawrence}, C.~R. and {Leach}, S. and {Leonardi}, R. and {Leroy}, C. and {Lilje}, P.~B. and {Linden-V{\o}rnle}, M. and {L{\'o}pez-Caniego}, M. and {Lubin}, P.~M. and {Mac{\'\i}as-P{\'e}rez}, J.~F. and {MacTavish}, C.~J. and {Maffei}, B. and {Maino}, D. and {Mandolesi}, N. and {Mann}, R. and {Maris}, M. and {Martin}, P. and {Mart{\'\i}nez-Gonz{\'a}lez}, E. and {Masi}, S. and {Matarrese}, S. and {Matthai}, F. and {Mazzotta}, P. and {McGehee}, P. and {Meinhold}, P.~R. and {Melchiorri}, A. and {Mendes}, L. and {Mennella}, A. and {Miville-Desch{\^e}nes}, M. -A. and {Moneti}, A. and {Montier}, L. and {Morgante}, G. and {Mortlock}, D. and {Munshi}, D. and {Murphy}, A. and {Naselsky}, P. and {Natoli}, P. and {Netterfield}, C.~B. and {N{\o}rgaard-Nielsen}, H.~U. and {Noviello}, F. and {Novikov}, D. and {Novikov}, I. and {O'Dwyer}, I.~J. and {Onishi}, T. and {Osborne}, S. and {Pajot}, F. and {Paladini}, R. and {Paradis}, D. and {Pasian}, F. and {Patanchon}, G. and {Perdereau}, O. and {Perotto}, L. and {Perrotta}, F. and {Piacentini}, F. and {Piat}, M. and {Plaszczynski}, S. and {Pointecouteau}, E. and {Polenta}, G. and {Ponthieu}, N. and {Poutanen}, T. and {Pr{\'e}zeau}, G. and {Prunet}, S. and {Puget}, J. -L. and {Reach}, W.~T. and {Reinecke}, M. and {Renault}, C. and {Ricciardi}, S. and {Riller}, T. and {Ristorcelli}, I. and {Rocha}, G. and {Rosset}, C. and {Rowan-Robinson}, M. and {Rubi{\~n}o-Mart{\'\i}n}, J.~A. and {Rusholme}, B. and {Sandri}, M. and {Santos}, D. and {Savini}, G. and {Scott}, D. and {Seiffert}, M.~D. and {Shellard}, P. and {Smoot}, G.~F. and {Starck}, J. -L. and {Stivoli}, F. and {Stolyarov}, V. and {Stompor}, R. and {Sudiwala}, R. and {Sygnet}, J. -F. and {Tauber}, J.~A. and {Terenzi}, L. and {Toffolatti}, L. and {Tomasi}, M. and {Torre}, J. -P. and {Tristram}, M. and {Tuovinen}, J. and {Umana}, G. and {Valenziano}, L. and {Vielva}, P.},
        title = "{Planck early results. XIX. All-sky temperature and dust optical depth from Planck and IRAS. Constraints on the ``dark gas'' in our Galaxy}",
      journal = {\aap},
         year = 2011,
        month = dec,
       volume = {536},
          eid = {A19},
        pages = {A19},
          doi = {10.1051/0004-6361/201116479},
archivePrefix = {arXiv},
       eprint = {1101.2029},
 primaryClass = {astro-ph.GA},
       adsurl = {https://ui.adsabs.harvard.edu/abs/2011A&A...536A..19P}
}

@ARTICLE{2010ApJ...716.1191W,
       author = {{Wolfire}, Mark G. and {Hollenbach}, David and {McKee}, Christopher F.},
        title = "{The Dark Molecular Gas}",
      journal = {\apj},
         year = 2010,
        month = jun,
       volume = {716},
       number = {2},
        pages = {1191-1207},
          doi = {10.1088/0004-637X/716/2/1191},
archivePrefix = {arXiv},
       eprint = {1004.5401},
 primaryClass = {astro-ph.GA},
       adsurl = {https://ui.adsabs.harvard.edu/abs/2010ApJ...716.1191W}
}

@ARTICLE{2019arXiv190206398M,
       author = {{Matsuo}, Hiroshi and {Shi}, Sheng-Cai and {Paine}, Scott and {Yao}, Qi-Jun and {Lin}, Zhen-Hui},
        title = "{Terahertz Atmospheric Windows for High Angular Resolution Terahertz Astronomy from Dome A}",
      journal = {arXiv e-prints},
         year = 2019,
        month = feb,
          eid = {arXiv:1902.06398},
        pages = {arXiv:1902.06398},
          doi = {10.48550/arXiv.1902.06398},
archivePrefix = {arXiv},
       eprint = {1902.06398},
 primaryClass = {astro-ph.IM},
       adsurl = {https://ui.adsabs.harvard.edu/abs/2019arXiv190206398M}
}

@INPROCEEDINGS{2008SPIE.7012E..49K,
       author = {{Kulesa}, C.~A. and {Walker}, C.~K. and {Schein}, M. and {Golish}, D. and {Tothill}, N. and {Siegel}, P. and {Weinreb}, S. and {Jones}, G. and {Bardin}, J. and {Jacobs}, K. and {Martin}, C.~L. and {Storey}, J. and {Ashley}, M. and {Lawrence}, J. and {Luong-Van}, D. and {Everett}, J. and {Wang}, L. and {Feng}, L. and {Zhu}, Z. and {Yan}, J. and {Yang}, J. and {Zhang}, X. -G. and {Cui}, X. and {Yuan}, X. and {Hu}, J. and {Xu}, Z. and {Jiang}, Z. and {Yang}, H. and {Li}, Y. and {Sun}, B. and {Qin}, W. and {Shang}, Z.},
        title = "{Pre-HEAT: submillimeter site testing and astronomical spectra from Dome A, Antarctica}",
    booktitle = {Ground-based and Airborne Telescopes II},
         year = 2008,
       editor = {{Stepp}, Larry M. and {Gilmozzi}, Roberto},
       series = {Society of Photo-Optical Instrumentation Engineers (SPIE) Conference Series},
       volume = {7012},
        month = jul,
          eid = {701249},
        pages = {701249},
          doi = {10.1117/12.789741},
       adsurl = {https://ui.adsabs.harvard.edu/abs/2008SPIE.7012E..49K}
}

@ARTICLE{2020RAA....20..168S,
       author = {{Shang}, Zhaohui},
        title = "{Astronomy from Dome A in Antarctica}",
      journal = {Research in Astronomy and Astrophysics},
         year = 2020,
        month = oct,
       volume = {20},
       number = {10},
          eid = {168},
        pages = {168},
          doi = {10.1088/1674-4527/20/10/168},
archivePrefix = {arXiv},
       eprint = {2010.04972},
 primaryClass = {astro-ph.IM},
       adsurl = {https://ui.adsabs.harvard.edu/abs/2020RAA....20..168S}
}

@ARTICLE{2023A&A...679L...5B,
       author = {{Bonne}, L. and {Kabanovic}, S. and {Schneider}, N. and {Zavagno}, A. and {Keilmann}, E. and {Simon}, R. and {Buchbender}, C. and {G{\"u}sten}, R. and {Jacob}, A.~M. and {Jacobs}, K. and {Kavak}, U. and {Polles}, F.~L. and {Tiwari}, M. and {Wyrowski}, F. and {Tielens}, A.~G.~G.~M.},
        title = "{The SOFIA FEEDBACK [CII] Legacy Survey: Rapid molecular cloud dispersal in RCW 79}",
      journal = {\aap},
         year = 2023,
        month = nov,
       volume = {679},
          eid = {L5},
        pages = {L5},
          doi = {10.1051/0004-6361/202347721},
archivePrefix = {arXiv},
       eprint = {2310.01657},
 primaryClass = {astro-ph.GA},
       adsurl = {https://ui.adsabs.harvard.edu/abs/2023A&A...679L...5B}
}

@ARTICLE{2015ApJ...806....7T,
       author = {{Torii}, K. and {Hasegawa}, K. and {Hattori}, Y. and {Sano}, H. and {Ohama}, A. and {Yamamoto}, H. and {Tachihara}, K. and {Soga}, S. and {Shimizu}, S. and {Okuda}, T. and {Mizuno}, N. and {Onishi}, T. and {Mizuno}, A. and {Fukui}, Y.},
        title = "{Cloud-Cloud Collision as a Trigger of the High-mass Star Formation: a Molecular Line Study in RCW120}",
      journal = {\apj},
         year = 2015,
        month = jun,
       volume = {806},
       number = {1},
          eid = {7},
        pages = {7},
          doi = {10.1088/0004-637X/806/1/7},
archivePrefix = {arXiv},
       eprint = {1503.00070},
 primaryClass = {astro-ph.GA},
       adsurl = {https://ui.adsabs.harvard.edu/abs/2015ApJ...806....7T}
}

@ARTICLE{2017A&A...600A..93F,
       author = {{Figueira}, M. and {Zavagno}, A. and {Deharveng}, L. and {Russeil}, D. and {Anderson}, L.~D. and {Men'shchikov}, A. and {Schneider}, N. and {Hill}, T. and {Motte}, F. and {M{\`e}ge}, P. and {LeLeu}, G. and {Roussel}, H. and {Bernard}, J. -P. and {Traficante}, A. and {Paradis}, D. and {Tig{\'e}}, J. and {Andr{\'e}}, P. and {Bontemps}, S. and {Abergel}, A.},
        title = "{Star formation towards the Galactic H II region RCW 120. Herschel observations of compact sources}",
      journal = {\aap},
         year = 2017,
        month = apr,
       volume = {600},
          eid = {A93},
        pages = {A93},
          doi = {10.1051/0004-6361/201629379},
archivePrefix = {arXiv},
       eprint = {1612.08862},
 primaryClass = {astro-ph.GA},
       adsurl = {https://ui.adsabs.harvard.edu/abs/2017A&A...600A..93F}
}

@article{research.0586,
author = {Jing Li  and Xianjin Deng  and Yangmei Li  and Jie Hu  and Wei Miao  and Changxing Lin  and Jun Jiang  and Shengcai Shi },
title = {Terahertz Science and Technology in Astronomy, Telecommunications, and Biophysics},
journal = {Research},
volume = {8},
number = {},
pages = {0586},
year = {2025},
doi = {10.34133/research.0586},
URL = {https://spj.science.org/doi/abs/10.34133/research.0586},
eprint = {https://spj.science.org/doi/pdf/10.34133/research.0586}}

@ARTICLE{1988ApJ...334..771V,
       author = {{van Dishoeck}, Ewine F. and {Black}, John H.},
        title = "{The Photodissociation and Chemistry of Interstellar CO}",
      journal = {\apj},
         year = 1988,
        month = nov,
       volume = {334},
        pages = {771},
          doi = {10.1086/166877},
       adsurl = {https://ui.adsabs.harvard.edu/abs/1988ApJ...334..771V}
}

@ARTICLE{2018PASJ...70S..45O,
       author = {{Ohama}, Akio and {Kohno}, Mikito and {Hasegawa}, Keisuke and {Torii}, Kazufumi and {Nishimura}, Atsushi and {Hattori}, Yusuke and {Hayakawa}, Takahiro and {Inoue}, Tsuyoshi and {Sano}, Hidetoshi and {Yamamoto}, Hiroaki and {Tachihara}, Kengo and {Fukui}, Yasuo},
        title = "{The formation of a Spitzer bubble RCW 79 triggered by a cloud-cloud collision}",
      journal = {\pasj},
         year = 2018,
        month = may,
       volume = {70},
          eid = {S45},
        pages = {S45},
          doi = {10.1093/pasj/psy025},
archivePrefix = {arXiv},
       eprint = {1709.02320},
 primaryClass = {astro-ph.GA},
       adsurl = {https://ui.adsabs.harvard.edu/abs/2018PASJ...70S..45O}
}

@ARTICLE{2022A&A...659A..36K,
       author = {{Kabanovic}, S. and {Schneider}, N. and {Ossenkopf-Okada}, V. and {Falasca}, F. and {G{\"u}sten}, R. and {Stutzki}, J. and {Simon}, R. and {Buchbender}, C. and {Anderson}, L. and {Bonne}, L. and {Guevara}, C. and {Higgins}, R. and {Koribalski}, B. and {Luisi}, M. and {Mertens}, M. and {Okada}, Y. and {R{\"o}llig}, M. and {Seifried}, D. and {Tiwari}, M. and {Wyrowski}, F. and {Zavagno}, A. and {Tielens}, A.~G.~G.~M.},
        title = "{Self-absorption in [C II], $^{12}$CO, and H I in RCW120. Building up a geometrical and physical model of the region}",
      journal = {\aap},
         year = 2022,
        month = mar,
       volume = {659},
          eid = {A36},
        pages = {A36},
          doi = {10.1051/0004-6361/202142575},
archivePrefix = {arXiv},
       eprint = {2112.11336},
 primaryClass = {astro-ph.GA},
       adsurl = {https://ui.adsabs.harvard.edu/abs/2022A&A...659A..36K}
}

@ARTICLE{2006A&A...446..171Z,
       author = {{Zavagno}, A. and {Deharveng}, L. and {Comer{\'o}n}, F. and {Brand}, J. and {Massi}, F. and {Caplan}, J. and {Russeil}, D.},
        title = "{Triggered massive-star formation on the borders of Galactic H II regions. II. Evidence for the collect and collapse process around RCW 79}",
      journal = {\aap},
         year = 2006,
        month = jan,
       volume = {446},
       number = {1},
        pages = {171-184},
          doi = {10.1051/0004-6361:20053952},
archivePrefix = {arXiv},
       eprint = {astro-ph/0509289},
 primaryClass = {astro-ph},
       adsurl = {https://ui.adsabs.harvard.edu/abs/2006A&A...446..171Z}
}

@ARTICLE{2007A&A...472..835Z,
       author = {{Zavagno}, A. and {Pomar{\`e}s}, M. and {Deharveng}, L. and {Hosokawa}, T. and {Russeil}, D. and {Caplan}, J.},
        title = "{Triggered star formation on the borders of the Galactic H ii region RCW 120}",
      journal = {\aap},
         year = 2007,
        month = sep,
       volume = {472},
       number = {3},
        pages = {835-846},
          doi = {10.1051/0004-6361:20077474},
archivePrefix = {arXiv},
       eprint = {0707.1185},
 primaryClass = {astro-ph},
       adsurl = {https://ui.adsabs.harvard.edu/abs/2007A&A...472..835Z}
}

@ARTICLE{2017A&A...602A..95L,
       author = {{Liu}, Hong-Li and {Figueira}, Miguel and {Zavagno}, Annie and {Hill}, Tracey and {Schneider}, Nicola and {Men'shchikov}, Alexander and {Russeil}, Delphine and {Motte}, Fr{\'e}d{\'e}rique and {Tig{\'e}}, J{\'e}r{\'e}my and {Deharveng}, Lise and {Anderson}, Loren D. and {Li}, Jin-Zeng and {Wu}, Yuefang and {Yuan}, Jing-Hua and {Huang}, Maohai},
        title = "{Herschel observations of the Galactic H II region RCW 79}",
      journal = {\aap},
         year = 2017,
        month = jun,
       volume = {602},
          eid = {A95},
        pages = {A95},
          doi = {10.1051/0004-6361/201629915},
archivePrefix = {arXiv},
       eprint = {1702.01924},
 primaryClass = {astro-ph.GA},
       adsurl = {https://ui.adsabs.harvard.edu/abs/2017A&A...602A..95L}
}

@ARTICLE{1999ApJ...527L..59I,
       author = {{Ikeda}, Masafumi and {Maezawa}, Hiroyuki and {Ito}, Tetsuya and {Saito}, Gaku and {Sekimoto}, Yutaro and {Yamamoto}, Satoshi and {Tatematsu}, Ken'ichi and {Arikawa}, Yuji and {Aso}, Yoshiyuki and {Noguchi}, Takashi and {Shi}, Sheng-Cai and {Miyazawa}, Keisuke and {Saito}, Shuji and {Ozeki}, Hiroyuki and {Fujiwara}, Hideo and {Ohishi}, Masatoshi and {Inatani}, Junji},
        title = "{Large-Scale Mapping Observations of the C I ($^{3}$P$_{1}$- $^{3}$P$_{0}$) and CO (J = 3-2) Lines toward the Orion A Molecular Cloud}",
      journal = {\apjl},
         year = 1999,
        month = dec,
       volume = {527},
       number = {1},
        pages = {L59-L62},
          doi = {10.1086/312395},
archivePrefix = {arXiv},
       eprint = {astro-ph/9912261},
 primaryClass = {astro-ph},
       adsurl = {https://ui.adsabs.harvard.edu/abs/1999ApJ...527L..59I}
}

@ARTICLE{2013ApJ...774L..20S,
       author = {{Shimajiri}, Yoshito and {Sakai}, Takeshi and {Tsukagoshi}, Takashi and {Kitamura}, Yoshimi and {Momose}, Munetake and {Saito}, Masao and {Oshima}, Tai and {Kohno}, Kotaro and {Kawabe}, Ryohei},
        title = "{Extensive [C I] Mapping toward the Orion-A Giant Molecular Cloud}",
      journal = {\apjl},
         year = 2013,
        month = sep,
       volume = {774},
       number = {2},
          eid = {L20},
        pages = {L20},
          doi = {10.1088/2041-8205/774/2/L20},
archivePrefix = {arXiv},
       eprint = {1308.1036},
 primaryClass = {astro-ph.GA},
       adsurl = {https://ui.adsabs.harvard.edu/abs/2013ApJ...774L..20S}
}

@ARTICLE{2020PASP..132j4301S,
       author = {{Schneider}, N. and {Simon}, R. and {Guevara}, C. and {Buchbender}, C. and {Higgins}, R.~D. and {Okada}, Y. and {Stutzki}, J. and {G{\"u}sten}, R. and {Anderson}, L.~D. and {Bally}, J. and {Beuther}, H. and {Bonne}, L. and {Bontemps}, S. and {Chambers}, E. and {Csengeri}, T. and {Graf}, U.~U. and {Gusdorf}, A. and {Jacobs}, K. and {Justen}, M. and {Kabanovic}, S. and {Karim}, R. and {Luisi}, M. and {Menten}, K. and {Mertens}, M. and {Mookerjea}, B. and {Ossenkopf-Okada}, V. and {Pabst}, C. and {Pound}, M.~W. and {Richter}, H. and {Reyes}, N. and {Ricken}, O. and {R{\"o}llig}, M. and {Russeil}, D. and {S{\'a}nchez-Monge}, {\'A}. and {Sandell}, G. and {Tiwari}, M. and {Wiesemeyer}, H. and {Wolfire}, M. and {Wyrowski}, F. and {Zavagno}, A. and {Tielens}, A.~G.~G.~M.},
        title = "{FEEDBACK: a SOFIA Legacy Program to Study Stellar Feedback in Regions of Massive Star Formation}",
      journal = {\pasp},
         year = 2020,
        month = oct,
       volume = {132},
       number = {1016},
          eid = {104301},
        pages = {104301},
          doi = {10.1088/1538-3873/aba840},
archivePrefix = {arXiv},
       eprint = {2009.08730},
 primaryClass = {astro-ph.GA},
       adsurl = {https://ui.adsabs.harvard.edu/abs/2020PASP..132j4301S}
}

@ARTICLE{2023A&A...670A..98Y,
       author = {{Yan}, Y.~T. and {Henkel}, C. and {Kobayashi}, C. and {Menten}, K.~M. and {Gong}, Y. and {Zhang}, J.~S. and {Yu}, H.~Z. and {Yang}, K. and {Xie}, J.~J. and {Wang}, Y.~X.},
        title = "{Direct measurements of carbon and sulfur isotope ratios in the Milky Way}",
      journal = {\aap},
         year = 2023,
        month = feb,
       volume = {670},
          eid = {A98},
        pages = {A98},
          doi = {10.1051/0004-6361/202244584},
archivePrefix = {arXiv},
       eprint = {2212.03252},
 primaryClass = {astro-ph.GA},
       adsurl = {https://ui.adsabs.harvard.edu/abs/2023A&A...670A..98Y}
}

@ARTICLE{1982ApJ...262..590F,
       author = {{Frerking}, M.~A. and {Langer}, W.~D. and {Wilson}, R.~W.},
        title = "{The relationship between carbon monoxide abundance and visual extinction in interstellar clouds.}",
      journal = {\apj},
         year = 1982,
        month = nov,
       volume = {262},
        pages = {590-605},
          doi = {10.1086/160451},
       adsurl = {https://ui.adsabs.harvard.edu/abs/1982ApJ...262..590F}
}

@ARTICLE{1978ApJ...224..132B,
       author = {{Bohlin}, R.~C. and {Savage}, B.~D. and {Drake}, J.~F.},
        title = "{A survey of interstellar H I from Lalpha absorption measurements. II.}",
      journal = {\apj},
         year = 1978,
        month = aug,
       volume = {224},
        pages = {132-142},
          doi = {10.1086/156357},
       adsurl = {https://ui.adsabs.harvard.edu/abs/1978ApJ...224..132B}
}

@ARTICLE{2006ApJ...649..268S,
       author = {{Sakai}, Takeshi and {Oka}, Tomoharu and {Yamamoto}, Satoshi},
        title = "{Atomic Carbon in the AFGL 333 Cloud}",
      journal = {\apj},
         year = 2006,
        month = sep,
       volume = {649},
       number = {1},
        pages = {268-279},
          doi = {10.1086/504861},
       adsurl = {https://ui.adsabs.harvard.edu/abs/2006ApJ...649..268S}
}

@ARTICLE{2023A&A...672A..36L,
       author = {{Liu}, Daizhong and {Schinnerer}, Eva and {Saito}, Toshiki and {Rosolowsky}, Erik and {Leroy}, Adam and {Usero}, Antonio and {Sandstrom}, Karin and {Klessen}, Ralf S. and {Glover}, Simon C.~O. and {Ao}, Yiping and {Be{\v{s}}li{\'c}}, Ivana and {Bigiel}, Frank and {Cao}, Yixian and {Chastenet}, J{\'e}r{\'e}my and {Chevance}, M{\'e}lanie and {Dale}, Daniel A. and {Gao}, Yu and {Hughes}, Annie and {Kreckel}, Kathryn and {Kruijssen}, J.~M. Diederik and {Pan}, Hsi-An and {Pety}, J{\'e}r{\^o}me and {Salak}, Dragan and {Santoro}, Francesco and {Schruba}, Andreas and {Sun}, Jiayi and {Teng}, Yu-Hsuan and {Williams}, Thomas},
        title = "{C I and CO in nearby spiral galaxies. I. Line ratio and abundance variations at {\ensuremath{\sim}}200 pc scales}",
      journal = {\aap},
         year = 2023,
        month = apr,
       volume = {672},
          eid = {A36},
        pages = {A36},
          doi = {10.1051/0004-6361/202244564},
archivePrefix = {arXiv},
       eprint = {2212.09661},
 primaryClass = {astro-ph.GA},
       adsurl = {https://ui.adsabs.harvard.edu/abs/2023A&A...672A..36L}
}

@ARTICLE{2015ApJ...803...37B,
       author = {{Bisbas}, Thomas G. and {Papadopoulos}, Padelis P. and {Viti}, Serena},
        title = "{Effective Destruction of CO by Cosmic Rays: Implications for Tracing H$_{2}$ Gas in the Universe}",
      journal = {\apj},
         year = 2015,
        month = apr,
       volume = {803},
       number = {1},
          eid = {37},
        pages = {37},
          doi = {10.1088/0004-637X/803/1/37},
archivePrefix = {arXiv},
       eprint = {1502.04198},
 primaryClass = {astro-ph.GA},
       adsurl = {https://ui.adsabs.harvard.edu/abs/2015ApJ...803...37B}
}

@ARTICLE{2004ApJ...612..921B,
       author = {{Bergin}, Edwin A. and {Hartmann}, Lee W. and {Raymond}, John C. and {Ballesteros-Paredes}, Javier},
        title = "{Molecular Cloud Formation behind Shock Waves}",
      journal = {\apj},
         year = 2004,
        month = sep,
       volume = {612},
       number = {2},
        pages = {921-939},
          doi = {10.1086/422578},
archivePrefix = {arXiv},
       eprint = {astro-ph/0405329},
 primaryClass = {astro-ph},
       adsurl = {https://ui.adsabs.harvard.edu/abs/2004ApJ...612..921B}
}

@ARTICLE{2015ApJ...811...13B,
       author = {{Burton}, Michael G. and {Ashley}, Michael C.~B. and {Braiding}, Catherine and {Freeman}, Matthew and {Kulesa}, Craig and {Wolfire}, Mark G. and {Hollenbach}, David J. and {Rowell}, Gavin and {Lau}, James},
        title = "{Extended Carbon Line Emission in the Galaxy: Searching for Dark Molecular Gas along the G328 Sightline}",
      journal = {\apj},
         year = 2015,
        month = sep,
       volume = {811},
       number = {1},
          eid = {13},
        pages = {13},
          doi = {10.1088/0004-637X/811/1/13},
archivePrefix = {arXiv},
       eprint = {1508.04828},
 primaryClass = {astro-ph.GA},
       adsurl = {https://ui.adsabs.harvard.edu/abs/2015ApJ...811...13B}
}

@ARTICLE{1997ApJ...480L..59S,
       author = {{Stark}, Antony A. and {Bolatto}, Alberto D. and {Chamberlin}, Richard A. and {Lane}, Adair P. and {Bania}, T.~M. and {Jackson}, James M. and {Lo}, K. -Y.},
        title = "{First Detection of 492 GHz [C I] Emission from the Large Magellanic Cloud}",
      journal = {\apjl},
         year = 1997,
        month = may,
       volume = {480},
       number = {1},
        pages = {L59-L62},
          doi = {10.1086/310618},
       adsurl = {https://ui.adsabs.harvard.edu/abs/1997ApJ...480L..59S}
}

@ARTICLE{2001ApJ...548..253O,
       author = {{Ojha}, Roopesh and {Stark}, Antony A. and {Hsieh}, Henry H. and {Lane}, Adair P. and {Chamberlin}, Richard A. and {Bania}, Thomas M. and {Bolatto}, Alberto D. and {Jackson}, James M. and {Wright}, Gregory A.},
        title = "{AST/RO Observations of Atomic Carbon near the Galactic Center}",
      journal = {\apj},
         year = 2001,
        month = feb,
       volume = {548},
       number = {1},
        pages = {253-257},
          doi = {10.1086/318693},
archivePrefix = {arXiv},
       eprint = {astro-ph/0008439},
 primaryClass = {astro-ph},
       adsurl = {https://ui.adsabs.harvard.edu/abs/2001ApJ...548..253O}
}

@ARTICLE{2001ApJ...553..274Z,
       author = {{Zhang}, Xiaolei and {Lee}, Youngung and {Bolatto}, Alberto and {Stark}, Antony A.},
        title = "{CO (J=4-->3) and [C I] Observations of the Carina Molecular Cloud Complex}",
      journal = {\apj},
         year = 2001,
        month = may,
       volume = {553},
       number = {1},
        pages = {274-287},
          doi = {10.1086/320628},
archivePrefix = {arXiv},
       eprint = {astro-ph/0101272},
 primaryClass = {astro-ph},
       adsurl = {https://ui.adsabs.harvard.edu/abs/2001ApJ...553..274Z}
}

@ARTICLE{2006A&A...454L..13G,
       author = {{G{\"u}sten}, R. and {Nyman}, L. {\r{A}}. and {Schilke}, P. and {Menten}, K. and {Cesarsky}, C. and {Booth}, R.},
        title = "{The Atacama Pathfinder EXperiment (APEX) - a new submillimeter facility for southern skies -}",
      journal = {\aap},
         year = 2006,
        month = aug,
       volume = {454},
       number = {2},
        pages = {L13-L16},
          doi = {10.1051/0004-6361:20065420},
       adsurl = {https://ui.adsabs.harvard.edu/abs/2006A&A...454L..13G}
}

@ARTICLE{2016PASJ...68...10I,
       author = {{Ishii}, Shun and {Seta}, Masumichi and {Nagai}, Makoto and {Miyamoto}, Yusuke and {Nakai}, Naomasa and {Nagasaki}, Taketo and {Arai}, Hitoshi and {Imada}, Hiroaki and {Miyagawa}, Naoki and {Maezawa}, Hiroyuki and {Maehashi}, Hideki and {Bronfman}, Leonardo and {Finger}, Ricardo},
        title = "{Large-scale CO (J = 4-3) mapping toward the Orion-A giant molecular cloud}",
      journal = {\pasj},
         year = 2016,
        month = feb,
       volume = {68},
       number = {1},
          eid = {10},
        pages = {10},
          doi = {10.1093/pasj/psv116},
archivePrefix = {arXiv},
       eprint = {1511.01940},
 primaryClass = {astro-ph.GA},
       adsurl = {https://ui.adsabs.harvard.edu/abs/2016PASJ...68...10I}
}

@ARTICLE{1999RvMP...71..173H,
       author = {{Hollenbach}, D.~J. and {Tielens}, A.~G.~G.~M.},
        title = "{Photodissociation regions in the interstellar medium of galaxies}",
      journal = {Reviews of Modern Physics},
         year = 1999,
        month = jan,
       volume = {71},
       number = {1},
        pages = {173-230},
          doi = {10.1103/RevModPhys.71.173},
       adsurl = {https://ui.adsabs.harvard.edu/abs/1999RvMP...71..173H}
}

@ARTICLE{2022ARA&A..60..247W,
       author = {{Wolfire}, Mark G. and {Vallini}, Livia and {Chevance}, M{\'e}lanie},
        title = "{Photodissociation and X-Ray-Dominated Regions}",
      journal = {\araa},
         year = 2022,
        month = aug,
       volume = {60},
        pages = {247-318},
          doi = {10.1146/annurev-astro-052920-010254},
archivePrefix = {arXiv},
       eprint = {2202.05867},
 primaryClass = {astro-ph.GA},
       adsurl = {https://ui.adsabs.harvard.edu/abs/2022ARA&A..60..247W}
}

@ARTICLE{2019Natur.565..618P,
       author = {{Pabst}, C. and {Higgins}, R. and {Goicoechea}, J.~R. and {Teyssier}, D. and {Berne}, O. and {Chambers}, E. and {Wolfire}, M. and {Suri}, S.~T. and {Guesten}, R. and {Stutzki}, J. and {Graf}, U.~U. and {Risacher}, C. and {Tielens}, A.~G.~G.~M.},
        title = "{Disruption of the Orion molecular core 1 by wind from the massive star {\ensuremath{\theta}}$^{1}$ Orionis C}",
      journal = {\nat},
         year = 2019,
        month = jan,
       volume = {565},
       number = {7741},
        pages = {618-621},
          doi = {10.1038/s41586-018-0844-1},
archivePrefix = {arXiv},
       eprint = {1901.04221},
 primaryClass = {astro-ph.GA},
       adsurl = {https://ui.adsabs.harvard.edu/abs/2019Natur.565..618P}
}

@ARTICLE{2021SciA....7.9511L,
       author = {{Luisi}, Matteo and {Anderson}, Loren D. and {Schneider}, Nicola and {Simon}, Robert and {Kabanovic}, Slawa and {G{\"u}sten}, Rolf and {Zavagno}, Annie and {Broos}, Patrick S. and {Buchbender}, Christof and {Guevara}, Cristian and {Jacobs}, Karl and {Justen}, Matthias and {Klein}, Bernd and {Linville}, Dylan and {R{\"o}llig}, Markus and {Russeil}, Delphine and {Stutzki}, J{\"u}rgen and {Tiwari}, Maitraiyee and {Townsley}, Leisa K. and {Tielens}, Alexander G.~G.~M.},
        title = "{Stellar feedback and triggered star formation in the prototypical bubble RCW 120}",
      journal = {Science Advances},
         year = 2021,
        month = apr,
       volume = {7},
       number = {15},
        pages = {eabe9511},
          doi = {10.1126/sciadv.abe9511},
archivePrefix = {arXiv},
       eprint = {2104.04568},
 primaryClass = {astro-ph.GA},
       adsurl = {https://ui.adsabs.harvard.edu/abs/2021SciA....7.9511L}
}

@ARTICLE{2023NatAs...7..546S,
       author = {{Schneider}, Nicola and {Bonne}, Lars and {Bontemps}, Sylvain and {Kabanovic}, Slawa and {Simon}, Robert and {Ossenkopf-Okada}, Volker and {Buchbender}, Christof and {Stutzki}, J{\"u}rgen and {Mertens}, Marc and {Ricken}, Oliver and {Csengeri}, Timea and {Tielens}, Alexander G.~G.~M.},
        title = "{Ionized carbon as a tracer of the assembly of interstellar clouds}",
      journal = {Nature Astronomy},
         year = 2023,
        month = may,
       volume = {7},
        pages = {546-556},
          doi = {10.1038/s41550-023-01901-5},
archivePrefix = {arXiv},
       eprint = {2302.09266},
 primaryClass = {astro-ph.GA},
       adsurl = {https://ui.adsabs.harvard.edu/abs/2023NatAs...7..546S}
}

@INPROCEEDINGS{2018SPIE10700E..1MS,
       author = {{Stacey}, G.~J. and {Aravena}, M. and {Basu}, K. and {Battaglia}, N. and {Beringue}, B. and {Bertoldi}, F. and {Bond}, J.~R. and {Breysse}, P. and {Bustos}, R. and {Chapman}, S. and {Chung}, D.~T. and {Cothard}, N. and {Erler}, J. and {Fich}, M. and {Foreman}, S. and {Gallardo}, P. and {Giovanelli}, R. and {Graf}, U.~U. and {Haynes}, M.~P. and {Herrera-Camus}, R. and {Herter}, T.~L. and {Hlo{\v{z}}ek}, R. and {Johnstone}, D. and {Keating}, L. and {Magnelli}, B. and {Meerburg}, D. and {Meyers}, J. and {Murray}, N. and {Niemack}, M. and {Nikola}, T. and {Nolta}, M. and {Parshley}, S.~C. and {Riechers}, D.~A. and {Schilke}, P. and {Scott}, D. and {Stein}, G. and {Stevens}, J. and {Stutzki}, J. and {Vavagiakis}, E.~M. and {Viero}, M.~P.},
        title = "{CCAT-Prime: science with an ultra-widefield submillimeter observatory on Cerro Chajnantor}",
    booktitle = {Ground-based and Airborne Telescopes VII},
         year = 2018,
       editor = {{Marshall}, Heather K. and {Spyromilio}, Jason},
       series = {Society of Photo-Optical Instrumentation Engineers (SPIE) Conference Series},
       volume = {10700},
        month = jul,
          eid = {107001M},
        pages = {107001M},
          doi = {10.1117/12.2314031},
archivePrefix = {arXiv},
       eprint = {1807.04354},
 primaryClass = {astro-ph.GA},
       adsurl = {https://ui.adsabs.harvard.edu/abs/2018SPIE10700E..1MS}
}

@ARTICLE{2020arXiv200807453S,
       author = {{Sridharan}, T.~K. and {Bialy}, Shmuel and {Blundell}, Raymond and {Burkhardt}, Andrew and {Dame}, Thomas and {Doeleman}, Sheperd and {Finkbeiner}, Douglas and {Goodman}, Alyssa and {Grimes}, Paul and {Imara}, Nia and {Johnson}, Michael and {Keating}, Garrett and {Lada}, Charles and {Le Gal}, Romane and {Myers}, Philip and {Narayan}, Ramesh and {Paine}, Scott and {Patel}, Nimesh and {Raymond}, Alexander and {Tong}, Edward and {Wilner}, David and {Zhang}, Qizhou and {Zucker}, Catherine},
        title = "{A Prospective ISRO-CfA Himalayan Sub-millimeter-wave Observatory Initiative}",
      journal = {arXiv e-prints},
         year = 2020,
        month = aug,
          eid = {arXiv:2008.07453},
        pages = {arXiv:2008.07453},
          doi = {10.48550/arXiv.2008.07453},
archivePrefix = {arXiv},
       eprint = {2008.07453},
 primaryClass = {astro-ph.IM},
       adsurl = {https://ui.adsabs.harvard.edu/abs/2020arXiv200807453S}
}

@ARTICLE{2010A&ARv..18..417B,
       author = {{Burton}, Michael G.},
        title = "{Astronomy in Antarctica}",
      journal = {\aapr},
         year = 2010,
        month = oct,
       volume = {18},
       number = {4},
        pages = {417-469},
          doi = {10.1007/s00159-010-0032-2},
archivePrefix = {arXiv},
       eprint = {1007.2225},
 primaryClass = {astro-ph.IM},
       adsurl = {https://ui.adsabs.harvard.edu/abs/2010A&ARv..18..417B}
}

@ARTICLE{2025A&A...697L...2K,
       author = {{Keilmann}, E. and {Dannhauer}, S. and {Kabanovic}, S. and {Schneider}, N. and {Ossenkopf-Okada}, V. and {Simon}, R. and {Bonne}, L. and {Goldsmith}, P.~F. and {G{\"u}sten}, R. and {Zavagno}, A. and {Stutzki}, J. and {Riechers}, D. and {R{\"o}llig}, M. and {Verbena}, J.~L. and {Tielens}, A.~G.~G.~M.},
        title = "{[C II]-deficit caused by self-absorption in an ionized carbon-filled bubble in RCW79}",
      journal = {\aap},
         year = 2025,
        month = may,
       volume = {697},
          eid = {L2},
        pages = {L2},
          doi = {10.1051/0004-6361/202453445},
archivePrefix = {arXiv},
       eprint = {2504.08976},
 primaryClass = {astro-ph.GA},
       adsurl = {https://ui.adsabs.harvard.edu/abs/2025A&A...697L...2K}
}

@ARTICLE{2019ApJ...870...32K,
       author = {{Kuhn}, Michael A. and {Hillenbrand}, Lynne A. and {Sills}, Alison and {Feigelson}, Eric D. and {Getman}, Konstantin V.},
        title = "{Kinematics in Young Star Clusters and Associations with Gaia DR2}",
      journal = {\apj},
         year = 2019,
        month = jan,
       volume = {870},
       number = {1},
          eid = {32},
        pages = {32},
          doi = {10.3847/1538-4357/aaef8c},
archivePrefix = {arXiv},
       eprint = {1807.02115},
 primaryClass = {astro-ph.GA},
       adsurl = {https://ui.adsabs.harvard.edu/abs/2019ApJ...870...32K}
}

@ARTICLE{2010PASP..122..490Y,
       author = {{Yang}, H. and {Kulesa}, C.~A. and {Walker}, C.~K. and {Tothill}, N.~F.~H. and {Yang}, J. and {Ashley}, M.~C.~B. and {Cui}, X. and {Feng}, L. and {Lawrence}, J.~S. and {Luong-Van}, D.~M. and {McCaughrean}, M.~J. and {Storey}, J.~W.~V. and {Wang}, L. and {Zhou}, X. and {Zhu}, Z.},
        title = "{Exceptional Terahertz Transparency and Stability above Dome A, Antarctica}",
      journal = {\pasp},
         year = 2010,
        month = apr,
       volume = {122},
       number = {890},
        pages = {490},
          doi = {10.1086/652276},
       adsurl = {https://ui.adsabs.harvard.edu/abs/2010PASP..122..490Y}
}

@ARTICLE{2016NatAs...1E...1S,
       author = {{Shi}, Sheng-Cai and {Paine}, Scott and {Yao}, Qi-Jun and {Lin}, Zhen-Hui and {Li}, Xin-Xing and {Duan}, Wen-Ying and {Matsuo}, Hiroshi and {Zhang}, Qizhou and {Yang}, Ji and {Ashley}, M.~C.~B. and {Shang}, Zhaohui and {Hu}, Zhong-Wen},
        title = "{Terahertz and far-infrared windows opened at Dome A in Antarctica}",
      journal = {Nature Astronomy},
         year = 2016,
        month = dec,
       volume = {1},
          eid = {0001},
        pages = {0001},
          doi = {10.1038/s41550-016-0001},
archivePrefix = {arXiv},
       eprint = {1609.06015},
 primaryClass = {astro-ph.IM},
       adsurl = {https://ui.adsabs.harvard.edu/abs/2016NatAs...1E...1S}
}

@ARTICLE{2017ApJ...848...64K,
       author = {{Kuo}, Chao-Lin},
        title = "{Assessments of Ali, Dome A, and Summit Camp for mm-wave Observations Using MERRA-2 Reanalysis}",
      journal = {\apj},
         year = 2017,
        month = oct,
       volume = {848},
       number = {1},
          eid = {64},
        pages = {64},
          doi = {10.3847/1538-4357/aa8b74},
archivePrefix = {arXiv},
       eprint = {1707.08400},
 primaryClass = {astro-ph.IM},
       adsurl = {https://ui.adsabs.harvard.edu/abs/2017ApJ...848...64K}
}

@ARTICLE{2007A&A...468..627V,
       author = {{van der Tak}, F.~F.~S. and {Black}, J.~H. and {Sch{\"o}ier}, F.~L. and {Jansen}, D.~J. and {van Dishoeck}, E.~F.},
        title = "{A computer program for fast non-LTE analysis of interstellar line spectra. With diagnostic plots to interpret observed line intensity ratios}",
      journal = {\aap},
         year = 2007,
        month = jun,
       volume = {468},
       number = {2},
        pages = {627-635},
          doi = {10.1051/0004-6361:20066820},
archivePrefix = {arXiv},
       eprint = {0704.0155},
 primaryClass = {astro-ph},
       adsurl = {https://ui.adsabs.harvard.edu/abs/2007A&A...468..627V}
}

@ARTICLE{2020Atoms...8...15V,
       author = {{van der Tak}, Floris F.~S. and {Lique}, Fran{\c{c}}ois and {Faure}, Alexandre and {Black}, John H. and {van Dishoeck}, Ewine F.},
        title = "{The Leiden Atomic and Molecular Database (LAMDA): Current Status, Recent Updates, and Future Plans}",
      journal = {Atoms},
         year = 2020,
        month = apr,
       volume = {8},
       number = {2},
          eid = {15},
        pages = {15},
          doi = {10.3390/atoms8020015},
archivePrefix = {arXiv},
       eprint = {2004.11230},
 primaryClass = {astro-ph.GA},
       adsurl = {https://ui.adsabs.harvard.edu/abs/2020Atoms...8...15V}
}

@ARTICLE{2006ApJ...649..816N,
       author = {{Neufeld}, David A. and {Melnick}, Gary J. and {Sonnentrucker}, Paule and {Bergin}, Edwin A. and {Green}, Joel D. and {Kim}, Kyoung Hee and {Watson}, Dan M. and {Forrest}, William J. and {Pipher}, Judith L.},
        title = "{Spitzer Observations of HH 54 and HH 7-11: Mapping the H$_{2}$ Ortho-to-Para Ratio in Shocked Molecular Gas}",
      journal = {\apj},
         year = 2006,
        month = oct,
       volume = {649},
       number = {2},
        pages = {816-835},
          doi = {10.1086/506604},
archivePrefix = {arXiv},
       eprint = {astro-ph/0606232},
 primaryClass = {astro-ph},
       adsurl = {https://ui.adsabs.harvard.edu/abs/2006ApJ...649..816N}
}

@ARTICLE{2013PASP..125..306F,
       author = {{Foreman-Mackey}, Daniel and {Hogg}, David W. and {Lang}, Dustin and {Goodman}, Jonathan},
        title = "{emcee: The MCMC Hammer}",
      journal = {\pasp},
         year = 2013,
        month = mar,
       volume = {125},
       number = {925},
        pages = {306},
          doi = {10.1086/670067},
archivePrefix = {arXiv},
       eprint = {1202.3665},
 primaryClass = {astro-ph.IM},
       adsurl = {https://ui.adsabs.harvard.edu/abs/2013PASP..125..306F}
}

@ARTICLE{2010CAMCS...5...65G,
       author = {{Goodman}, Jonathan and {Weare}, Jonathan},
        title = "{Ensemble samplers with affine invariance}",
      journal = {Communications in Applied Mathematics and Computational Science},
         year = 2010,
        month = jan,
       volume = {5},
       number = {1},
        pages = {65-80},
          doi = {10.2140/camcos.2010.5.65},
       adsurl = {https://ui.adsabs.harvard.edu/abs/2010CAMCS...5...65G}
}

@ARTICLE{2013A&A...549A..85R,
       author = {{R{\"o}llig}, M. and {Szczerba}, R. and {Ossenkopf}, V. and {Gl{\"u}ck}, C.},
        title = "{Full SED fitting with the KOSMA-{\ensuremath{\tau}} PDR code. I. Dust modelling}",
      journal = {\aap},
         year = 2013,
        month = jan,
       volume = {549},
          eid = {A85},
        pages = {A85},
          doi = {10.1051/0004-6361/201118190},
archivePrefix = {arXiv},
       eprint = {1211.3546},
 primaryClass = {astro-ph.IM},
       adsurl = {https://ui.adsabs.harvard.edu/abs/2013A&A...549A..85R}
}

@ARTICLE{2022A&A...664A..67R,
       author = {{R{\"o}llig}, M. and {Ossenkopf-Okada}, V.},
        title = "{The KOSMA-{\ensuremath{\tau}} PDR model. I. Recent updates to the numerical model of photo-dissociated regions}",
      journal = {\aap},
         year = 2022,
        month = aug,
       volume = {664},
          eid = {A67},
        pages = {A67},
          doi = {10.1051/0004-6361/202141854},
archivePrefix = {arXiv},
       eprint = {2205.04233},
 primaryClass = {astro-ph.GA},
       adsurl = {https://ui.adsabs.harvard.edu/abs/2022A&A...664A..67R}
}

@ARTICLE{2023AJ....165...25P,
       author = {{Pound}, Marc W. and {Wolfire}, Mark G.},
        title = "{The PhotoDissociation Region Toolbox: Software and Models for Astrophysical Analysis}",
      journal = {\aj},
         year = 2023,
        month = jan,
       volume = {165},
       number = {1},
          eid = {25},
        pages = {25},
          doi = {10.3847/1538-3881/ac9b1f},
archivePrefix = {arXiv},
       eprint = {2210.08062},
 primaryClass = {astro-ph.IM},
       adsurl = {https://ui.adsabs.harvard.edu/abs/2023AJ....165...25P}
}
\bibliographystyle{sciencemag}

%
%
%
%
%
%


\section*{Acknowledgments}
We thank Xinmiao Jin, Bin Ma, Chao Chen, and all members of the 41st CHINARE for their strong support in carrying out the observations at Dome~A. The MeerKAT telescope is operated by the South African Radio Astronomy Observatory, which is a facility of the National Research Foundation, an agency of the Department of Science and Innovation. 
\paragraph*{Funding:}
This work was supported by the Strategic Priority Research Program of the Chinese Academy of Sciences (Grant No. XDB0800000) and the National Key Research and Development Program of China (Grant No. 2023YFA1608200). We also acknowledge support from the National Natural Science Foundation of China (NSFC) under Grants No. 12222308 and 12333013.
\paragraph*{Author contributions:}

Writing – original draft: Y.G., W.Z., W.M., J.Liu, Z.W., K.Z., B.L.L., X.M.L., S.M.X., Y.J.L., X.J.D., J.J., L.L., L.G., Y.Z., and J.Li \\
Writing - review \& editing: Y.G., J.Q.Z., Y.R., Y.L.Z., D.Z.L., Y.P.A., Q.J.Y., W.Z., W.M., Z.H.L., D.L., K.M.Z., J.Liu, Z.W., J.D.J., K.Z., F.W., J.P.L., B.L.L., X.Z., Z.H.L., J.M.W., H.Q.H., X.M.L., S.M.X., J.Q., Y.J.L., J.T.L., X.J.D., J.J., L.L., L.G., P.J., Y.Z., C.G.S., S.N., R.Q.M., S.C.S., J.Li \\
Data curation: Y.G., J.Q.Z., Y.R., Y.L.Z., D.Z.L., W.Z., W.M., J.Liu, Z.W., K.Z., F.W., X.Z., X.M.L., S.M.X., Y.J.L., L.L., L.G., Y.Z., S.N., and J.Li \\
Formal analysis: Y.G., Y.R., D.Z.L., W.Z., W.M., J.Liu, Z.W., K.Z., F.W., J.P.L., B.L.L., X.Z., X.M.L., S.M.X., Y.J.L., L.L., L.G., Y.Z., S.N., and J.Li \\
Visualization: Y.G., J.Q.Z., W.Z., W.M., J.Liu, Z.W., K.Z., F.W., B.L.L., X.Z., X.M.L., S.M.X., Y.J.L., J.T.L., L.L., L.G., Y.Z., S.N., and J.Li \\
Investigation: Y.G., J.Q.Z., Y.R., Y.L.Z., D.Z.L., Q.J.Y., W.Z., W.M., D.L., J.Liu, Z.W., J.D.J., K.Z., F.W., B.L.L., X.Z., X.M.L., S.M.X., Y.J.L., J.T.L., J.J., L.L., L.G., T.J., P.J., Y.Z., S.N., and J.Li \\
Methodology: Y.G., J.Q.Z., Y.R., D.Z.L., Q.J.Y., W.Z., W.M., W.Y.D., D.L., J.Liu, Z.W., K.Z., F.W., J.P.L., B.L.L., H.Q.H., X.M.L., S.M.X., J.T.L., X.J.D., J.J., L.L., L.G., Y.Z., S.N., S.C.S., and J.Li\\
Resources: Y.G., J.Q.Z., Y.R., D.Z.L., Q.J.Y., W.Z., W.M., Z.H.L., W.Y.D., K.M.Z., J.Liu, Z.W., J.D.J., K.Z., F.W., B.L.L., H.Q.H., X.M.L., S.M.X., Y.J.L., J.T.L., X.J.D., L.L., L.G., T.J., P.J., Y.Z., S.N., S.C.S., and J.Li \\
Software: Y.G., J.Q.Z., Y.R., D.Z.L., Q.J.Y., W.Z., W.M., Z.H.L., W.Y.D., J.Liu, K.Z., F.W., X.M.L., S.M.X., L.L., L.G., Y.Z., and J.Li \\
Conceptualization: Y.G., J.Q.Z., Y.R., D.Z.L., Q.J.Y., W.Z., W.M., D.L., J.Liu, Z.W., K.Z., F.W., X.M.L., S.M.X., J.T.L., X.J.D., J.J., L.L., L.G., P.J., Y.Z., R.Q.M., S.C.S., and J.Li\\
Validation: Y.G., J.Q.Z., Y.R., D.Z.L., Q.J.Y., W.Z., W.M., Z.H.L., D.L., J.Liu, Z.W., K.Z., F.W., J.P.L., B.L.L., X.Z., J.M.W., H.Q.H., X.M.L., S.M.X., J.Q., Y.J.L., J.T.L., L.L., L.G., Y.Z., C.G.S., S.N., and J.Li \\
Supervision: R.Q.M., S.C.S., and J.Li \\
Project administration: P.J., S.C.S., and J.Li \\
Funding acquisition: D.Z.L., C.G.S., R.Q.M., S.C.S., and J.Li \\

\paragraph*{Competing interests:}
There are no competing interests to declare.


\paragraph*{Data and materials availability:}
The reduced ATE60 CO ($4-3$) and [CI] data cubes are available on Zenodo at \url{https://doi.org/10.5281/zenodo.17410999}.
The $^{12}$CO ($1-0$), $^{13}$CO ($1-0$), and C$^{18}$O ($1-0$) data from the Three-mm Ultimate Mopra Milky Way Survey are available at \url{https://gemelli.spacescience.org/~pbarnes/research/thrumms/rbank/}.
The APEX $^{12}$CO ($3-2$) and $^{13}$CO ($3-2$) data are available at the CDS archive ( \url{http://cdsarc.u-strasbg.fr/viz-bin/cat/J/A+A/659/A36}).
The [CII] data are provided at the NASA/IPAC Infrared science archive (\url{https://irsa.ipac.caltech.edu/applications/sofia}).
The 1.3 GHz radio continuum data are provided at the SARAO Data Archive
(\url{https://archive-gw-1.kat.ac.za/public/repository/10.48479/3wfd-e270/index.html}).
The GLIMPSE 8~$\mu$m and MIPSGAL 24~$\mu$m data are available at the NASA/IPAC infrared science archive (\url{https://irsa.ipac.caltech.edu/Missions/spitzer.html}).
The H$_{2}$ column density maps derived from the Herschel Hi-GAL data are provided at \url{http://www.astro.cardiff.ac.uk/research/ViaLactea/}. The codes used for the ATE telescopes are publicly available via Zenodo (\url{https://doi.org/10.5281/zenodo.17410999}).

\subsection*{Supplementary materials}
Supplementary Text\\
Figs. S1 to S7\\
Tables S1 to S2\\


\newpage


\renewcommand{\thefigure}{S\arabic{figure}}
\renewcommand{\thetable}{S\arabic{table}}
\renewcommand{\theequation}{S\arabic{equation}}
\renewcommand{\thepage}{S\arabic{page}}
\setcounter{figure}{0}
\setcounter{table}{0}
\setcounter{equation}{0}
\setcounter{page}{1} 


\begin{center}
\section*{Supplementary Materials for\\ \scititle}

Yan Gong$^{1\dagger}$,
Jiaqiang Zhong$^{1,3\dagger}$,
Yuan Ren$^{1,3}$, 
Yilong Zhang$^{1,3}$, 
Daizhong Liu$^{1}$, \\
Yiping Ao$^{1,3}$, 
Qijun Yao$^{1}$, 
Wen Zhang$^{1,3}$, 
Wei Miao$^{1,3}$, 
Zhenhui Lin$^{1,3}$,  \\
Wenying Duan$^{1,2}$, 
Dong Liu$^{1}$, 
Kangmin Zhou$^{1}$, 
Jie Liu$^{1}$, 
Zheng Wang$^{1}$, 
Junda Jin$^{1}$, \\
Kun Zhang$^{1}$, 
Feng Wu$^{1}$, 
Jinpeng Li$^{1}$, 
Boliang Liu$^{1}$, 
Xuan Zhang$^{1,3}$, 
Zhengheng Luo$^{1,2}$, \\
Jiameng Wang$^{1,2}$, 
Huiqian Hao$^{4}$, 
Xingming Lu$^{4}$, 
Shaoming Xie$^{4}$, 
Jia Quan$^{5}$, \\
Yanjie Liu$^{5}$, 
Jingtao Liang$^{5}$, 
Xianjin Deng$^{6,7}$, 
Jun Jiang$^{6,7}$, 
Li Li$^{6,7}$, \\
Liang Guo$^{8}$, 
Tuo Ji$^{9}$, 
Peng Jiang$^{9}$, 
Yi Zhang$^{10}$, 
Chenggang Shu$^{10}$, \\
Sudeep Neupane$^{11}$, 
Ruiqing Mao$^{1}$, 
Shengcai Shi$^{1}$, 
Jing Li$^{1\ast}$ \\
    \small$^{1}$ Purple Mountain Observatory and Key Laboratory of Radio Astronomy, Chinese Academy of Sciences, 
    \and 
    \small Nanjing 210008, China 
    \and
	\small$^{2}$ School of Astronomy and Space Sciences, University of Science and Technology of China, Hefei 230026, China \and
    \small$^{3}$ State Key Laboratory of Radio Astronomy and Technology, National Astronomical Observatories, \and \small Chinese Academy of Sciences, Beijing 100101, China \and 
    \small$^{4}$ The 54th Research Institute of China Electronics Technology Group Corporation, Shijiazhuang 050081, China \and 
    \small$^{5}$ Technical Institute of Physics and Chemistry, Chinese Academy of Sciences, Beijing 100190, China \and 
    \small$^{6}$ Microsystem and Terahertz Research Center, China Academy of Engineering Physics, \and \small Chengdu, Sichuan 610200, China \and 
    \small$^{7}$ Institute of Electronic Engineering, China Academy of Engineering Physics, Mianyang, Sichuan 621999, China \and 
    \small$^{8}$ Changchun Institute of Optics, Fine Mechanics and Physics, Chinese Academy of Sciences, Changchun 130033, China \and 
    \small$^{9}$ Polar Research Institute of China; Key Laboratory for Polar Science, MNR, Shanghai 200136, China \and 
    \small$^{10}$ Shanghai Key Lab for Astrophysics, Shanghai Normal University, 100 Guilin Road, Shanghai 200234, China \and 
    \small$^{11}$ Max-Planck-Institut f{\"u}r Radioastronomie, Auf dem H{\"u}gel 69, D-53121 Bonn, Germany \and 
	\small$^\ast$Corresponding author. \quad Email: lijing@pmo.ac.cn    \and
	\small$^\dagger$These authors contributed equally to this work.

\end{center}
\normalsize
\subsubsection*{This PDF file includes:}
Supplementary Text\\
Figures S1 to S7\\
Tables S1 to S2\\


\newpage

\section*{Supplementary Text}
\subsection*{Archival Data}

\noindent \textbf{$\bullet$ [CII] 158~$\mu$m data.} We retrieve the [CII] 158~$\mu$m data of RCW~120 and RCW~79 from the SOFIA legacy program FEEDBACK \cite{2020PASP..132j4301S}. These [CII] observations were obtained using the upGREAT heterodyne spectrometer onboard SOFIA. The archival data have a HPBW of 14\rlap{.}\arcsec1 and a channel width is 0.2~\kms. The absolute flux calibration uncertainties were assumed to be 10\%\,for the [CII] data.

\noindent \textbf{$\bullet$ Low-$J$ CO data.} Both RCW~120 and RCW~79 have been extensively mapped in low-$J$ CO transitions. For both regions, the $^{12}$CO ($1-0$) and $^{13}$CO ($1-0$) data are taken from the data release 6 products of the Three-mm Ultimate Mopra Milky Way Survey\cite{2025ApJS..280...31B} with an HPBW of 72\arcsec. For RCW~120, additional $^{12}$CO ($3-2$) and $^{13}$CO ($3-2$) observations were obtained with APEX \cite{2022A&A...659A..36K}, providing a higher angular resolution with an HPBW of 20\arcsec. The absolute flux calibration uncertainties were assumed to be 10\%\,for the low-$J$ CO data.

\noindent \textbf{$\bullet$ Radio continuum data.} The 1.3~GHz radio continuum image of RCW~120 and RCW~79 were taken from the first data release (DR1) of the SARAO MeerKAT Galactic Plane Survey \cite{2024MNRAS.531..649G}, providing an angular resolution of 8\arcsec.

\noindent \textbf{$\bullet$ Infrared data.} The Spitzer 8~$\mu$m and 24~$\mu$m data were taken from the Galactic Legacy Infrared Mid-Plane Survey Extraordinaire (GLIMPSE\cite{2003PASP..115..953B}) and the Multiband Imaging Photometer for Spitzer Galactic Plane Survey (MIPSGAL\cite{2009PASP..121...76C}). Their angular resolutions are 1\rlap{.}\arcsec9 and 6\arcsec, respectively. 

\noindent \textbf{$\bullet$ Hi-GAL based H$_{2}$ column density maps.} The H$_{2}$ column density maps were generated by applying the point process mapping (PPMAP) technique\cite{2015MNRAS.454.4282M} to multi-wavelength Herschel imaging data, enabling the construction of high-resolution, dust temperature–resolved column density distributions across the Galactic plane\cite{2017MNRAS.471.2730M}. The Hi-GAL based H$_{2}$ column density maps achieve an angular resolution of 12\arcsec.

\subsection*{Source Selection}\label{sec.target}
In the SOFIA FEEDBACK survey\cite{2020PASP..132j4301S}, RCW~79 and RCW~120 stand out as the only two sources exhibiting prominent, well-defined ring-like morphologies in their PDRs (see Figure~\ref{fig:3col}). This striking structure provides compelling evidence that the surrounding gas has been profoundly reshaped by the feedback from massive stars. Both regions exhibit clear interactions between molecular clouds and H{\scriptsize II} regions, with previous studies suggesting that triggered star formation occurs in their peripheries \cite{2006A&A...446..171Z, 2007A&A...472..835Z, 2017A&A...602A..95L}. Intense UV radiation from massive stars likely drives photodissociation at the edges of these H{\scriptsize II} regions, making them ideal sites for studying the transition between different carbon phases. While both regions have been extensively studied in CO and [CII] emission\cite{2023A&A...679L...5B,2015ApJ...806....7T,2021SciA....7.9511L,2017A&A...600A..93F,2018PASJ...70S..45O,2022A&A...659A..36K,2025A&A...697L...2K}, [CI] observations are absent. Our ATE60 [CI] observations can address this crucial gap, enabling a more comprehensive study of the carbon cycle in these environments. RCW~79 and RCW~120 are located at a distance of $\sim$4~kpc\cite{2023A&A...679L...5B} and $\sim$1.7~kpc\cite{2019ApJ...870...32K}, respectively. Their large angular sizes of $\gtrsim 10^{\prime}$ allow for spatially resolved observations with ATE60. Situated in the circumpolar sky as seen from Dome~A, both sources are accessible for long-duration observations, making them ideal targets for this study.

\subsection*{Non-LTE Analysis}
To enhance the signal-to-noise ratios, all datasets were first convolved to a common HPBW of 6\arcmin. We selected four bright positions for a detailed analysis of their physical properties, and the corresponding spectra are presented in Figure~\ref{fig:spectral}. A single-component Gaussian profile was fitted to the spectral data to derive the observed parameters of most lines, whereas a two-component Gaussian profile was used to fit the [CII] spectra. The derived parameters are summarized in Table~\ref{tab:sample-info}.

In order to derive the physical properties of the selected regions, we use the non-local thermodynamic equilibrium (non-LTE) radiative
transfer code RADEX\footnote{\url{https://home.strw.leidenuniv.nl/~moldata/radex.html}} for the calculations\cite{2007A&A...468..627V}. In our calculations, we assume that CO, $^{13}$CO, and C$^{0}$ transitions trace gas with the same gas temperature and H$_{2}$ number density, so their transitions can be simultaneously modeled. The information on energy levels, statistical weights, Einstein A-coefficients and collisional rate coefficients of CO, $^{13}$CO, and C$^{0}$ transitions is directly taken from the Leiden Atomic and Molecular Database (LAMDA\footnote{\url{https://home.strw.leidenuniv.nl/~moldata/}})\cite{2020Atoms...8...15V}. 
During our modeling, we used the classic large-velocity gradient (LVG) approximation to estimate the escape probability. Based on previous studies\cite{2006ApJ...649..816N}, a constant H$_{2}$ ortho-to-para ratio of 0.25 is assumed in this study. The line widths are taken to be the fitted values of respective transitions (see Table~\ref{tab:sample-info}).

To reduce the number of the free parameters, we assume that the [$^{12}$CO/$^{13}$CO] abundance ratio approximates the [$^{12}$C/$^{13}$C] isotope ratio. The Galactocentric distance, $R_{\rm gc}$, was calculated using the expression
\begin{equation}\label{f.gc}
R_{\rm gc}=\sqrt{R_{\rm gc,\; \odot}^{2}+d^{2}-2R_{\rm gc,\; \odot}d{\rm cos}(l)} \;,
\end{equation}
where $R_{\rm gc, \odot}=8.15$ kpc is the Galactocentric distance of the Sun \cite{2019ApJ...885..131R}, $d$ is the heliocentric distance of the source, and $l$ is the Galactic longitude. For RCW~79 and RCW~120, we adopted a heliocentric distance of $\sim$4~kpc\cite{2023A&A...679L...5B} and $\sim$1.7~kpc\cite{2019ApJ...870...32K}, respectively, which yields Galactocentric distances of $\sim$6.5 kpc for both regions. Assuming that the [$^{12}$C/$^{13}$C] isotope ratio follows the gradient $^{12}$C/$^{13}$C$=4.77R_{\rm gc}+20.76$\cite{2023A&A...670A..98Y}, we used a representative [$^{12}$CO/$^{13}$CO] abundance ratio of 50 in this work.   

We performed parameter estimation using the \textit{emcee} package\cite{2013PASP..125..306F}, which implements the affine-invariant ensemble sampler\cite{2010CAMCS...5...65G} for the Markov chain Monte Carlo (MCMC) approach. This approach allowed us to explore the posterior probability distributions of the modeling parameters. Uniform priors were adopted for the kinetic temperature ($T_{\rm K}$), H$_{2}$ number density ($n_{\rm H_2}$), $^{13}$CO column density [log($N_{\rm ^{13}CO}$)], and the C$^{0}$-to-$^{13}$CO abundance ratio $N_{\rm C^0}/N_{\rm ^{13}CO}$. These parameters were allowed to vary within the following ranges: $T_{\rm K}= 5 - 300$~K, $n_{\rm H_2}$ = $10^{2} - 10^{7}$~cm$^{-3}$, $N_{\rm ^{13}CO}$ = $10^{13} - 10^{19}$ cm$^{-2}$, and $N_{\rm C^0}/N_{\rm ^{13}CO} = 0.1 - 100$.

The posterior distributions were computed as the product of the prior probability distributions and the likelihood function. We adopted a Gaussian likelihood of the form $\mathcal{L} \propto \exp(-\chi^{2}/2)$, where
\begin{equation}\label{f.chi}
\chi^{2} = \sum_{i} \left( \frac{I_{{\rm obs},i} - I_{{\rm mod},i}}{\sigma_{i}} \right)^{2} \;,
\end{equation}
with $I_{{\rm obs},i}$, $I_{{\rm mod},i}$, and $\sigma_{i}$ representing the observed, modeled integrated intensities, and the corresponding 1$\sigma$ observational uncertainties, respectively. For RCW~120, the fit included six transitions: CO ($1-0$), $^{13}$CO ($1-0$), CO ($3-2$), $^{13}$CO ($3-2$), CO ($4-3$), and [CI] ($^{3}P_{1}-^{3}P_{0}$). For RCW~79, four lines were used: CO (1–0), $^{13}$CO (1–0), CO (4–3), and [CI] ($^{3}P_{1} - ^{3}P_{0}$). The MCMC chains were run with 10 walkers and 6000 steps following an initial burn-in phase, ensuring convergence of the sampling. The final parameter estimates and their 1$\sigma$ uncertainties were derived from the 16th and 84th percentiles of the posterior distributions.

\begin{figure*}[!htbp]
\includegraphics[width=0.95\linewidth]{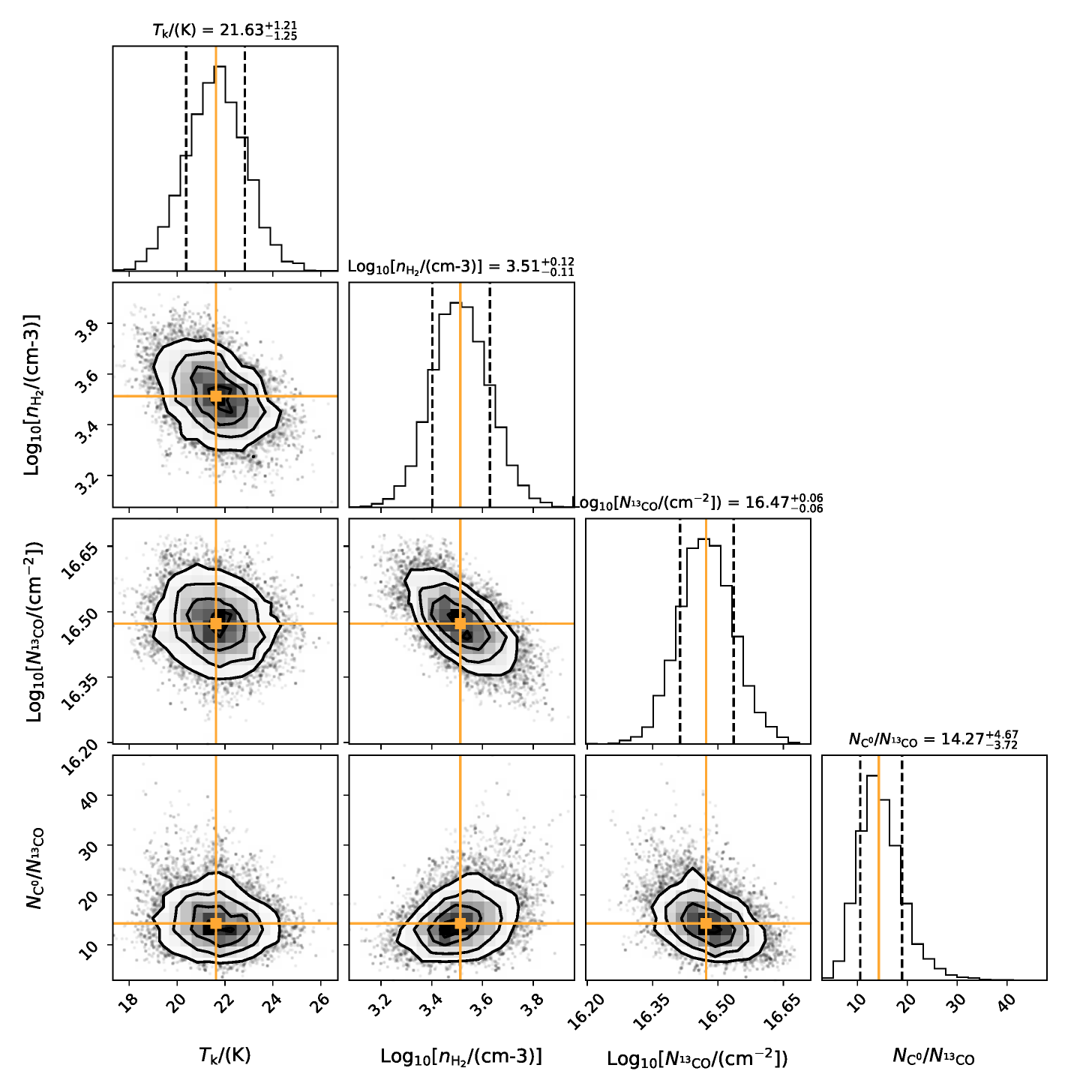}
\caption{Posterior probability distributions of $T_{\rm K}$, $n_{\rm H_2}$, log($N_{\rm ^{13}CO}$), and $N_{\rm C^0}/N_{\rm ^{13}CO}$ for RCW~120A from the RADEX models, with the maximum posterior possibility point in the parameter space highlighted by orange lines and points. Contours represent the 0.5, 1.0, 1.5, and 2.0$\sigma$ confidence intervals. The vertical dashed lines represent the 1$\sigma$ dispersion.}
\label{fig:radex}
\end{figure*}

Figure~\ref{fig:radex} presents a representative example of the modeling results for RCW~120A, where all physical parameters are well constrained. In contrast, $n_{\rm H_2}$ is not well constrained for RCW~79A and RCW~79B due to the limited number of detected transitions. To improve the reliability of the results for these two sources, we restrict $n_{\rm H_2}$ to the range of $10^{3}-10^{5}$~cm$^{-3}$ in the modeling. While $n_{\rm H_2}$ remains poorly constrained and carries larger uncertainties, the other parameters are robustly determined. A summary of the derived parameters for all four regions is presented in Table~\ref{tab:phy-info}.

\subsection*{PDR Models}
As demonstrated by previous studies, the intensity ratios between the transitions of C$^{+}$, C$^{0}$, and CO can be used to infer FUV radiation field strengths and gas densities\cite{2013A&A...549A..85R,2022A&A...664A..67R,2023AJ....165...25P}. Here, we adopted both homogeneous and clumpy PDR models available through the PDR Toolbox\footnote{\url{https://dustem.astro.umd.edu/}}\cite{2023AJ....165...25P}. Following previous studies\cite{2023AJ....165...25P}, we used the PDR Toolbox to fit selected observed line ratios that are effective tracers of local physical conditions. 

Various geometries have been proposed to model PDR structures. In this study, we primarily adopted two suites of PDR models: the Wolfire–Kaufman model and the KOSMA-$\tau$ model\cite{2013A&A...549A..85R,2022A&A...664A..67R}. The former assumes a plane-parallel geometry and corresponds to the ``Wolfire–Kaufman 2020 (wk2020)" implementation, while the latter is based on an ensemble of spherical clumps, either with a mass spectrum (referred to as ``clumpy") or a single clump mass (``non-clumpy"), as implemented in the KOSMA-$\tau$ model.
In all models, we assume typical solar metallicity ($z=1$). In the KOSMA-$\tau$ models, the clump mass is another free parameter. We use 1~$M_{\odot}$ as a fiducial case for the clump mass, and test the variation of the clump mass in the following calculations.  

\begin{figure*}[!htbp]
\includegraphics[width=0.95\linewidth]{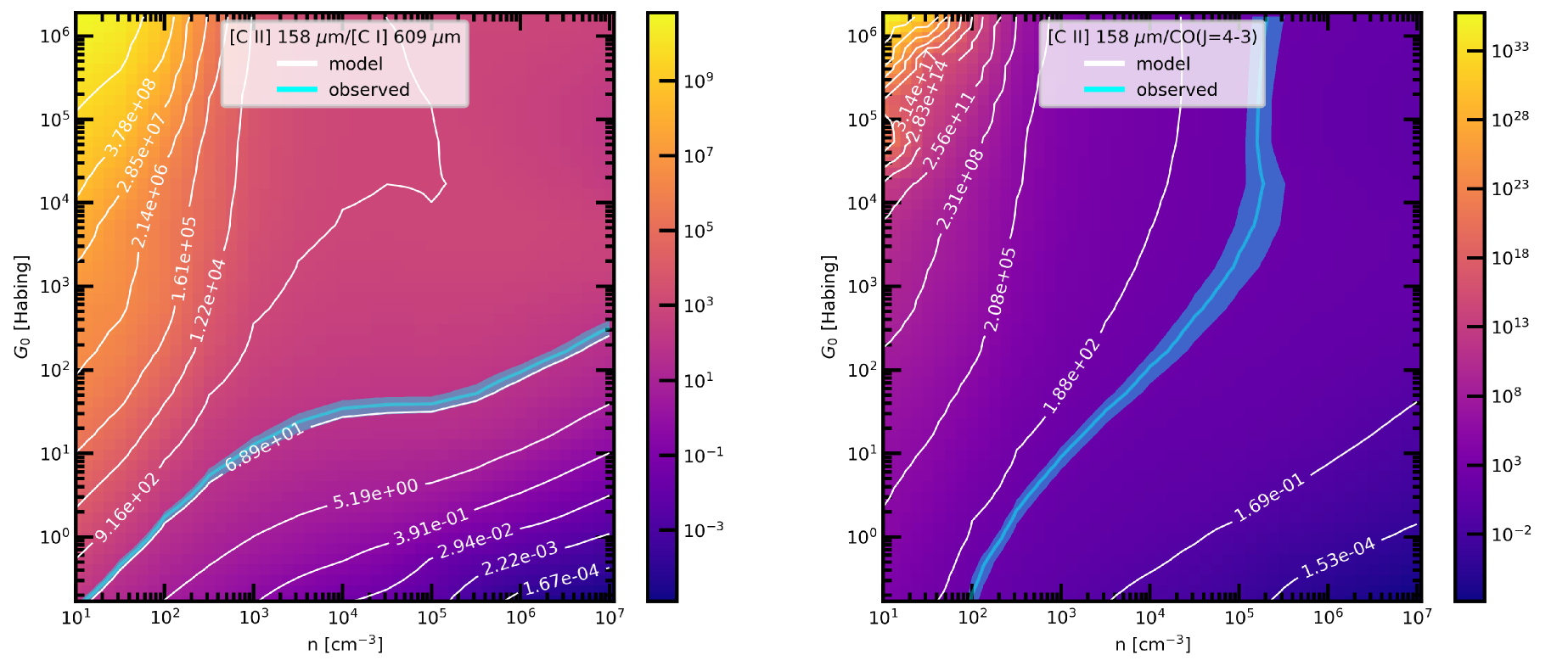}
\caption{Predictions of the observed intensity ratios as a function of $G_0$ and $n$ for RCW~120A, based on the clumpy PDR model. The solid lines represent the observed values, while the shaded regions indicate the $1\sigma$ errors.}
\label{fig:pdr}
\end{figure*}

Using RCW~120A as a representative case, we examine the observed line ratios as a function of the local radiation field strength and gas density in Figure~\ref{fig:pdr}. These comparisons demonstrate that multi-line observations can effectively constrain both parameters. To refine the model constraints, we incorporate archival low-$J$ transitions of CO and $^{13}$CO (see Section~\textit{Archival Data} in the supplementary materials). For the [CII] line, we considered only the narrow velocity component in the fitting, as the broad component is not detected in CO, $^{13}$CO, or [CI] lines and likely originates from a physically distinct gas component. Although self-absorption is present in the [CII] spectra of RCW~120A and RCW~120B (see Figure~\ref{fig:spectral}), Gaussian fitting still provides a reasonable estimate of the total flux.

\begin{figure*}[!htbp]
\includegraphics[width=0.95\linewidth]{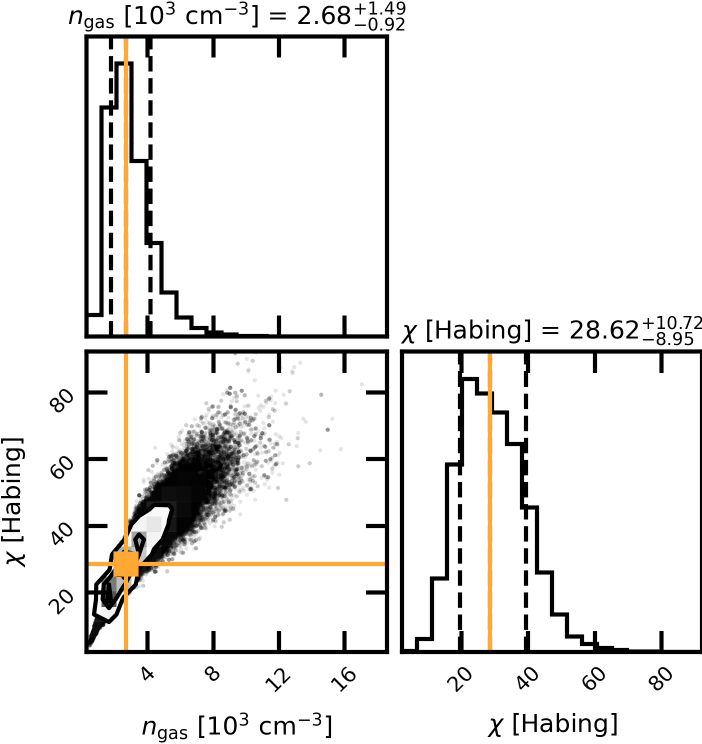}
\caption{Posterior probability distributions of gas density and radiation field
strength for RCW~120A from the KOSMA-$\tau$ clumpy PDR model, with the maximum posterior possibility point in the parameter space highlighted by orange lines and points. Contours represent the 0.5, 1.0, 1.5, and 2.0$\sigma$ confidence intervals. The vertical dashed lines represent the 1$\sigma$ dispersion.}
\label{fig:mcmc}
\end{figure*}

Figure~\ref{fig:mcmc} shows the modeling results of RCW~120A using the KOSMA-$\tau$ clumpy PDR framework, assuming a maximum clump mass of 1~$M_{\odot}$. Parameter estimation is also performed with the MCMC method, similar to that described in Section~\textit{Non-LTE Analysis} of the supplementary materials. The same procedure is applied to other PDR models, and the derived radiation field strengths and gas densities across the selected regions are summarized in Table~\ref{tab:phy-info}.

In the KOSMA-$\tau$ models, the clump mass might bias the inferred physical conditions. To assess this effect, we explore models with maximum clump masses of 0.1, 1, 10, and 100~$M_{\odot}$. In the clumpy PDR model, the results remain largely insensitive to variations in clump mass, indicating robustness against this parameter. In contrast, the non-clumpy version of the KOSMA-$\tau$ model shows a strong dependence on clump mass. Increasing the clump mass systematically yields higher estimates for both gas density and incident radiation field strength.

We further compare the results of these PDR models with our non-LTE analysis (see Section~\textit{Non-LTE Analysis} in the supplementary materials). In the case of a plane-parallel PDR, reproducing the observed line intensities requires very high gas densities and FUV radiation fields. However, such high gas densities are inconsistent with those derived from the non-LTE analysis, indicating that the plane-parallel scenario does not provide a satisfactory explanation for the observations. In contrast, the clumpy PDR model can reproduce the observed line properties with significantly lower gas densities and FUV field strengths. Taking the uncertainties into account, these values are in good agreement with the non-LTE results (see Section~\textit{Non-LTE Analysis} in the supplementary materials). Our findings thus favor a clumpy PDR structure over a homogeneous plane-parallel configuration in explaining the observed spectral characteristics. 

\begin{figure*}[!htbp]
\includegraphics[width=1.0\linewidth]{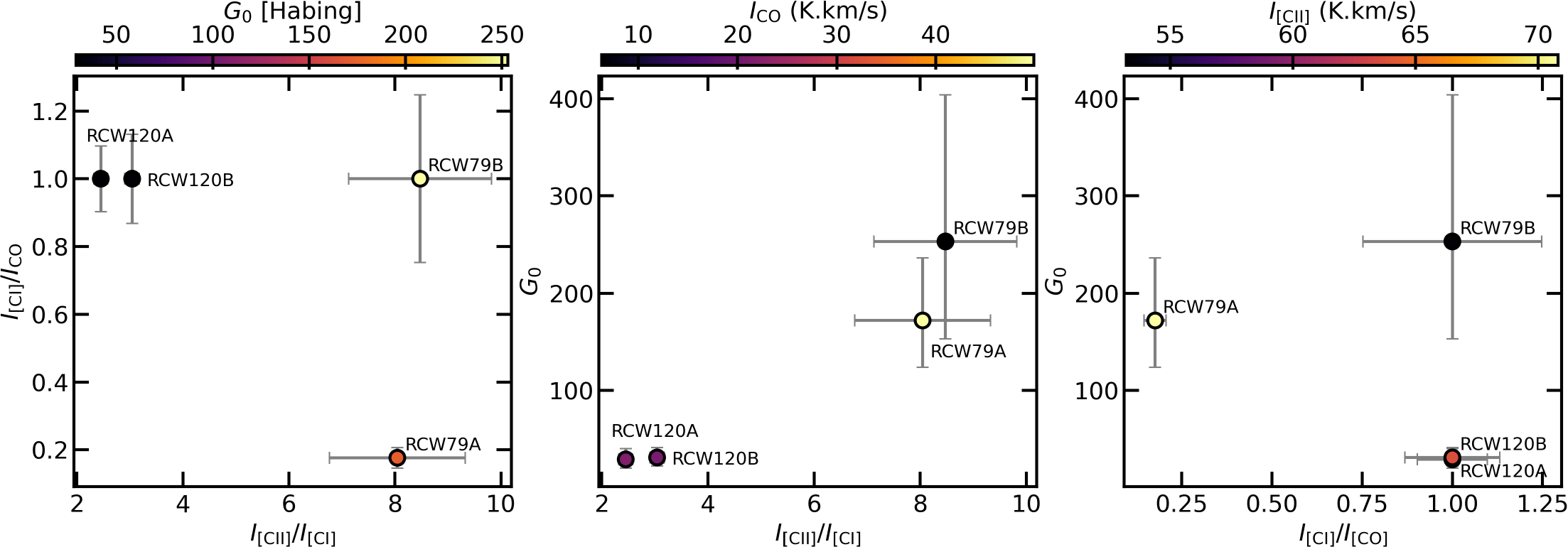}
\caption{Relationship between line ratios and radiation field strength. \textit{Left:} Observed integrated intensity ratio of [CI] to CO ($4-3$) as a function of the [CII]/[CI] ratio, with points colored by radiation field strength. \textit{Middle:} [CI]/CO ($4-3$) ratio versus radiation field strength, colored by CO ($4-3$) integrated intensity. \textit{Right:} Radiation field strength plotted against the [CI]/CO ($4-3$) ratio, with colors indicating [CII] integrated intensity.}
\label{fig:lineratio}
\end{figure*}

The relationships between line ratios and radiation field strengths are investigated in Figure~\ref{fig:lineratio}. The results show that the radiation field strength exhibits a strong correlation with the [CII]/[CI] integrated intensity ratio, but appears largely insensitive to the [CI]/CO ($4-3$) integrated intensity ratio. This highlights [CII] emission as an effective tracer of the radiation field strength. Notably, the C$^{0}$/CO abundance ratios appear to be comparable in RCW~79 and RCW~120 (see \ref{tab:phy-info}), despite a difference in radiation field strength of more than a factor of three between the two regions (see Table~\ref{tab:phy-info}). Given that C$^{0}$ and CO can coexist across a broad range of visual extinctions in clumpy PDRs, both species can remain similarly abundant in well-mixed zones. Alternatively, the transition layers between CO and C$^{0}$ may be similar across different radiation fields, or their contributions to the observed C$^{0}$/CO abundance ratios may be negligible. Together, these effects might account for the similar C$^{0}$/CO abundance ratios observed across environments with different radiation field strengths.

\begin{figure*}[!htbp]
\includegraphics[width=0.95\linewidth]{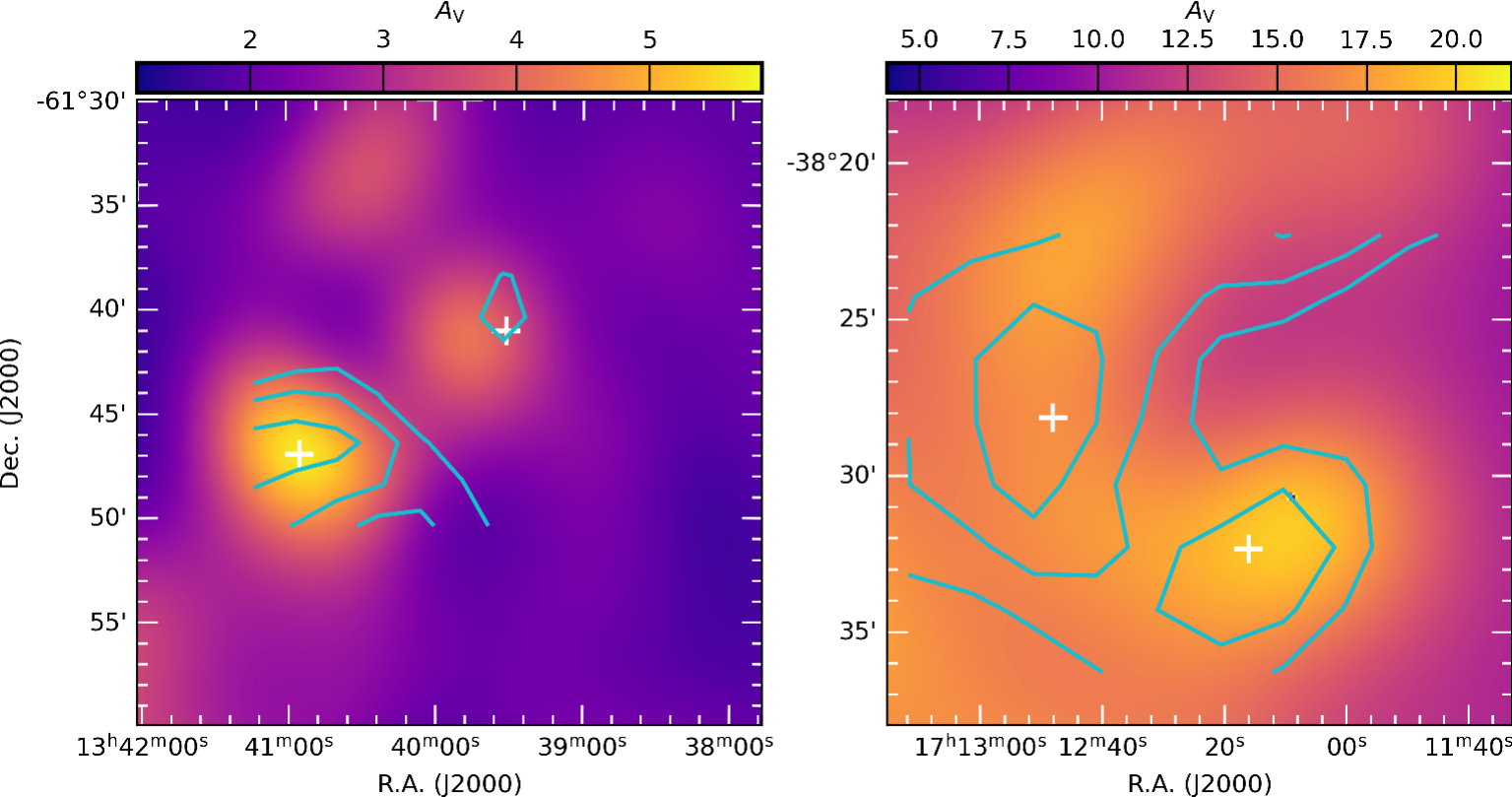}
\caption{Extinction maps of RCW~79 and RCW~120. The overlaid [CI] integrated intensity contours are identical to those shown in Figure~\ref{fig:3col}. The white pluses mark the positions of the four selected regions.}
\label{fig:av}
\end{figure*}

\subsection*{Extinction Map}
In addition to the visual extinction estimates presented in Section~\textit{Main text}, we also derive visual extinctions using Hi-GAL–based H$_{2}$ column density maps (see \textit{Archival Data}). To match the angular resolution of our observations, the H$_{2}$ column density maps were convolved to a beam size of 6\arcmin. Using the established relation between visual extinction and H$_{2}$ column density\cite{1978ApJ...224..132B}, we determine the extinction distributions for RCW~79 and RCW~120, as shown in Figure~\ref{fig:av}. 
The visual extinctions are within the range of 1.5--5.7 and 4.9--21.5 for RCW~79 and RCW~120, respectively. This independent method yields visual extinctions of approximately 6, 4, 17, and 19 toward RCW~79A, RCW~79B, RCW~120A, and RCW~120B, respectively.

\subsection*{Flux Calibration}
\begin{figure}[!htbp]
\includegraphics[width=.95\linewidth]{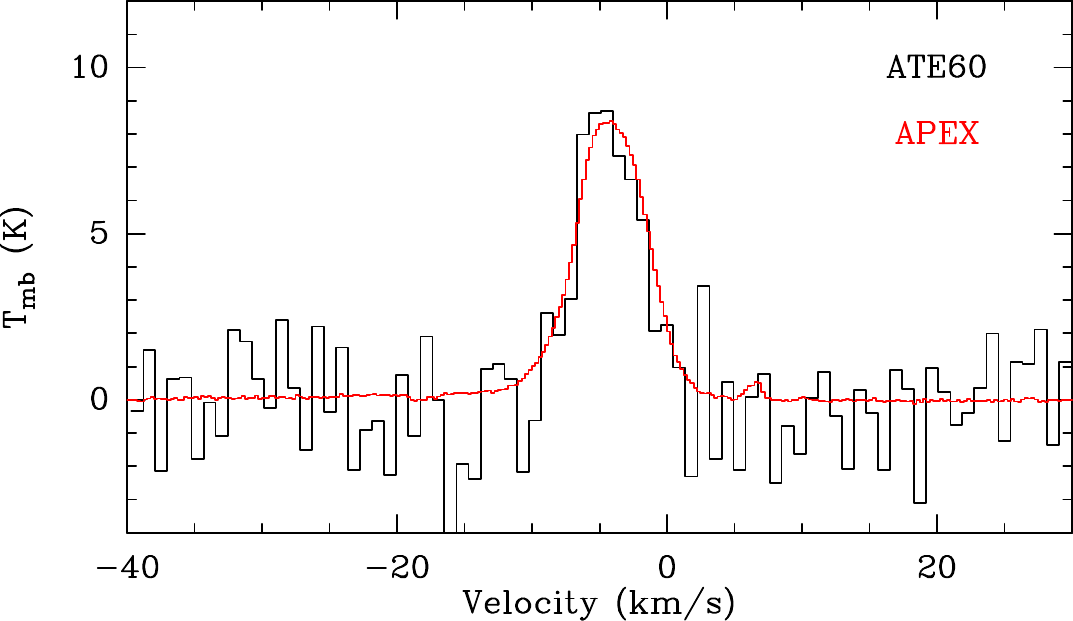}
\caption{\textbf{Comparison of [CI] ($^{3}P_{1}-^{3}P_{0}$) spectra of NGC~6334I obtained with ATE60 and APEX.} Spectra from ATE60 and APEX are shown in black and red, respectively. Main beam efficiencies of 35\% and 60\% were assumed for the ATE60 and APEX data, respectively, yielding good consistency between the two datasets.}
\label{fig:tmb}
\end{figure}

Absolute flux calibration was performed by comparing the integrated intensity of [CI] ($^{3}P_{1}-^{3}P_{0}$) from NGC~6334I obtained with ATE60 and APEX. The APEX data were observed under the large-scale [CI] mapping project of NGC~6334 (Project ID: M9516C\_0109, PI: Sudeep Neupane). The APEX data were convolved to the ATE60's angular resolution of 240\arcsec, and were calibrated with a main beam efficiency of 60\%\footnote{\url{https://www.apex-telescope.org/telescope/efficiency/?yearBy=2024}}. This comparison indicates a main beam efficiency of $\sim$35\%\,for ATE60 (see Figure~\ref{fig:tmb}), which was subsequently used to establish the main beam brightness temperature scale of all ATE60 data. The velocity was calibrated with respect to the local standard of rest (LSR), and the excellent agreement between the ATE60 and APEX data confirms the accuracy of the velocity calibration (see Figure~\ref{fig:tmb}). The absolute flux calibration uncertainties were assumed to be 20\%. 

\subsection*{Pointing}
\begin{figure*}[!htbp]
\includegraphics[width=0.95\linewidth]{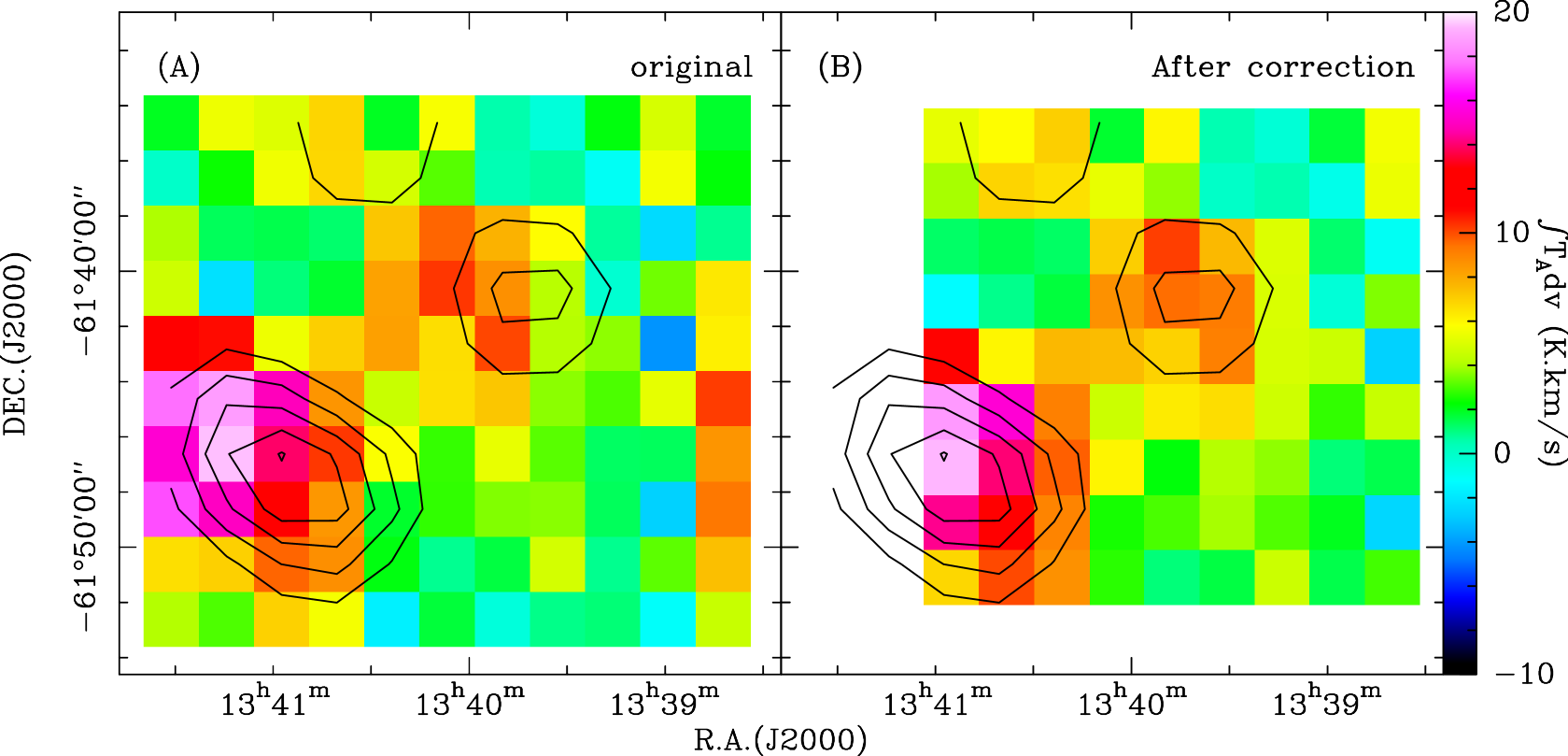}
\caption{\textbf{Distribution of CO ($4-3$) and $^{13}$CO ($1-0$) in RCW~79.}
Integrated-intensity map of CO ($4-3$) overlaid with $^{13}$CO ($1-0$) contours, before (panel A) and after (panel B) pointing correction. The Mopra $^{13}$CO ($1-0$) data have been convolved to 270\arcsec, matching the beam size of our CO ($4-3$) data. The $^{13}$CO ($1-0$) contours start at 8~K~\kms\,and increase by 2~K~\kms. Both maps are integrated over the velocity range of $-$40 to $-$30 km s$^{-1}$. In panel (b), the image has been cropped to exclude artifacts introduced by the re-gridding process.}
\label{fig:pointing}
\end{figure*}

The pointing model was first established toward Canopus using an accompanying off-axis optical telescope. However, the accuracy is not high enough with respect to the ATE60's HPBWs. Based on previous studies\cite{2016PASJ...68...10I}, the distributions of $^{13}$CO ($1-0$) and CO ($4-3$) are similar on scales of $\gtrsim$1~pc. To assess the residual pointing errors after the correction from the optical telescope, we compared Mopra $^{13}$CO ($1-0$) data with our ATE60 CO ($4-3$) data. The $^{13}$CO ($1-0$) data were convolved to have an angular resolution of 270\arcsec\,for comparison. The comparison revealed an accuracy of $\lesssim 3^{\prime}$ in right ascension and declination (see Figure~\ref{fig:pointing}A). To further improve the pointing accuracy, we refined the pointing by cross-matching bright, compact sources in CO ($4-3$) with $^{13}$CO ($1-0$), assuming a common spatial origin for these lines. The peak position in the CO ($4-3$) map was used to estimate the offset in azimuth and elevation, which was assumed to be identical across the entire map. Incorporating the offset, the observed coordinates were then re-calculated. The updated coordinates were no longer regularly spaced, prompting re-gridding of the data for subsequent analysis. As shown in Figure~\ref{fig:pointing}B, the regridded ATE60 CO ($4-3$) image exhibits a good spatial agreement with the Mopra $^{13}$CO ($1-0$) image, demonstrating the effectiveness of our pointing correction method. Ultimately, the pointing error was estimated to be $\lesssim$1$^{\prime}$ after correction.

\begin{table}[!hbtp]
\footnotesize
\centering
\caption{{\bf Observed properties of CO, [CI], and [CII] transitions toward the selected positions in RCW~79 and RCW~120 at a common angular resolution of 6\arcmin.} Source: source name; $\alpha_{\rm J2000}$: right ascension; $\delta_{\rm J2000}$: declination; Transition: observed spectral transition; $\varv_{\rm lsr}$: velocity centroid; $\Delta \varv$: FWHM line width; $T_{\rm p}$: peak main beam temperature; $\int T_{\rm mb} {\rm d}\varv$: Integrated intensity.}
\label{tab:sample-info}
\begin{threeparttable}
\begin{tabular}{cccccccc}
\hline
Source            & ($\alpha_{\rm J2000}$) & ($\delta_{\rm J2000}$) & Transition & $\varv_{\rm lsr}$ & $\Delta \varv$  & $T_{\rm p}$      & $\int T_{\rm mb}{\rm d}\varv$     \\
                  & ($^{\rm h}$:$^{\rm m}$:$^{\rm s}$) & ($^{\circ}$:$^{\prime}$:$^{\prime\prime}$) &  & (\kms)  & (\kms)          & (K)              &  (K~\kms)  \\
\hline
RCW~79A & 13:40:55 & $-$61:46:59 & CO ($1-0$) & $-$46.7$\pm$0.1 & 4.8$\pm$0.1 & 6.8$\pm$0.3  & 35.0$\pm$0.6 \\
&   &   & $^{13}$CO ($1-0$) & $-$46.7$\pm$0.1 & 3.6$\pm$0.1  & 3.1$\pm$0.1 & 11.9$\pm$0.2   \\
&   &   & CO ($4-3$)        & $-$46.5$\pm$0.2 & 5.6$\pm$0.5  & 8.4$\pm$0.3 & 49.9$\pm$3.4   \\
&   &   &[CI] ($^{3}P_{1}-^{3}P_{0}$)& $-$46.2$\pm$0.3 & 4.1$\pm$0.7 & 2.0$\pm$0.3 & 8.8$\pm$1.4\\   
&   &   &[CII] ($^{2}P_{3/2}-^{2}P_{1/2}$)\tnote{a}& $-$47.5$\pm$0.1 & 18.2$\pm$0.2 & 2.3$\pm$0.2 & 44.3$\pm$0.5   \\
&   &   &                                 & $-$46.4$\pm$0.1 & 4.8$\pm$0.1 & 5.2$\pm$0.2 & 26.5$\pm$0.2   \\
\hline
RCW~79B & 13:39:31 & $-$61:41:04 & CO ($1-0$) & $-$46.7$\pm$0.1 & 3.8$\pm$0.2  & 3.9$\pm$0.3 & 16.5$\pm$0.6 \\
         &   &   & $^{13}$CO ($1-0$) & $-$46.8$\pm$0.1 & 4.0$\pm$0.3  & 1.4$\pm$0.2 & 6.1$\pm$0.4 \\
         &   &   & CO ($4-3$)        & $-$46.6$\pm$0.1 & 5.1$\pm$0.4  & 4.0$\pm$0.5 & 21.8$\pm$1.2 \\
         &   &   &[CI] ($^{3}P_{1}-^{3}P_{0}$)& $-$46.2$\pm$0.4 & 4.5$\pm$0.8 & 1.3$\pm$0.3 & 6.3$\pm$1.0        \\   
         &   &   &[CII] ($^{2}P_{3/2}-^{2}P_{1/2}$)\tnote{a}& $-$46.7$\pm$0.1 & 16.7$\pm$0.3 & 1.9$\pm$0.1  & 34.1$\pm$0.5   \\
         &   &   &                                 & $-$47.0$\pm$0.1 & 5.0$\pm$0.1 & 3.6$\pm$0.1  & 19.3$\pm$0.2   \\
\hline
RCW~120A & 17:12:48 & $-$38:28:10 & CO ($1-0$) & $-$8.4$\pm$0.1 & 6.5$\pm$0.1 & 11.7$\pm$0.8      & 80.8$\pm$0.7   \\
         &   &   & $^{13}$CO ($1-0$) & $-$7.8$\pm$0.1 & 4.2$\pm$0.1 & 5.8$\pm$0.8 & 26.0$\pm$0.4 \\
         &   &   & CO ($3-2$) & $-$7.8$\pm$0.1 & 6.3$\pm$0.1   & 9.4$\pm$0.3       &  62.8$\pm$0.3  \\
         &   &   & $^{13}$CO ($3-2$) & $-$7.3$\pm$0.1 & 3.3$\pm$0.1 & 4.3$\pm$0.1  & 14.8$\pm$0.1  \\
         &   &   & CO ($4-3$)        & $-$7.6$\pm$0.1 & 5.1$\pm$0.2 & 8.8$\pm$0.6  & 47.9$\pm$1.8   \\
         &   &   &[CI] ($^{3}P_{1}-^{3}P_{0}$)& $-$7.5$\pm$0.1 & 4.8$\pm$0.3     & 4.3$\pm$0.5  & 21.7$\pm$1.1 \\   
         &   &   &[CII] ($^{2}P_{3/2}-^{2}P_{1/2}$)\tnote{a}& $-$8.5$\pm$0.2 & 17.5$\pm$0.6 & 1.1$\pm$0.2  & 20.9$\pm$0.8   \\
         &   &     & & $-$6.8$\pm$0.1 & 6.0$\pm$0.1 & 5.0$\pm$0.3  & 32.3$\pm$0.2   \\         
\hline
RCW~120B  & 17:12:16 & $-$38:32:22 & CO ($1-0$) & $-$8.3$\pm$0.1 & 6.4$\pm$0.1 & 10.7$\pm$0.7 &  72.3$\pm$0.4  \\
         &   &   & $^{13}$CO ($1-0$) & $-$7.9$\pm$0.1 & 4.0$\pm$0.1 & 5.2$\pm$0.3 & 22.5$\pm$0.3 \\
         &   &   & CO ($3-2$) & $-$7.8$\pm$0.1 & 6.7$\pm$0.5 & 8.4$\pm$0.5  & 59.3$\pm$0.2 \\
         &   &   & $^{13}$CO ($3-2$) & $-$7.3$\pm$0.1 & 3.3$\pm$0.1 & 4.3$\pm$0.1 & 14.9$\pm$0.1 \\
         &   &   & CO ($4-3$)        & $-$7.7$\pm$0.1 & 5.8$\pm$0.3 & 9.5$\pm$0.7 & 58.6$\pm$2.6 \\
         &   &   &[CI] ($^{3}P_{1}-^{3}P_{0}$)& $-$7.9$\pm$0.1 & 5.2$\pm$0.3 & 3.7$\pm$0.3 & 20.3$\pm$0.9  \\   
         &   &   &[CII] ($^{2}P_{3/2}-^{2}P_{1/2}$)\tnote{a}& $-$9.7$\pm$0.1 & 13.6$\pm$0.3 & 2.6$\pm$0.2 & 37.1$\pm$0.8 \\
         &   &   &                                 & $-$6.9$\pm$0.1 & 5.2$\pm$0.1 & 4.8$\pm$0.4 & 26.5$\pm$0.2 \\
\hline
\end{tabular}
\begin{tablenotes}
        \item[a] [CII] spectra were fitted by assuming two velocity components, consisting of a broad-line-width and a narrow-line-width component. 
\end{tablenotes}
\end{threeparttable}
\normalsize
\end{table}

\begin{table*}[!hbt]
       \footnotesize
\centering
\caption{{\bf Physical properties of the selected positions derived from the carbon's three primary phases.} $T_{\rm K}$: gas kinetic temperature; $n_{\rm H_{2}}$: H$_{2}$ number density; $N_{\rm ^{13}CO}$: $^{13}$CO column density; $N_{\rm C^{0}}$: C$^{0}$ column density; $N_{\rm C^{0}}/N_{\rm ^{13}CO}$: the column density ratio between C$^{0}$ and $^{13}$CO; $N_{\rm C^{0}}/N_{\rm CO}$: the column density ratio between C$^{0}$ and CO; $n_{\rm gas}$: gas density; $G_{0}$ radiation field;}
\label{tab:phy-info}
\begin{threeparttable}
\begin{tabular}{ccccccc}
\hline
            & \multicolumn{6}{c}{Non-LTE}      \\
        \cmidrule(lr){2-7}        
Source   & $T_{\rm K}$ & log$_{10}(n_{\rm H_{2}})$  & log$_{10}(N_{\rm ^{13}CO})$ & $N_{\rm C^{0}}$  & $N_{\rm C^{0}}/N_{\rm ^{13}CO}$ & $N_{\rm C^{0}}/N_{\rm CO}$\tnote{a} \\
         &  (K) &  (cm$^{-3}$)  & (cm$^{-2}$)  &  ($\times 10^{17}$~cm$^{-2}$)   &   & \\
\hline
RCW~79A  & $14.8^{+1.2}_{-1.2}$  & $4.3^{+0.5}_{-0.6}$  & $16.2^{+0.1}_{-0.1}$ & 2.4$^{+1.3}_{-1.0}$ & $15.2^{+7.5}_{-5.3}$ & $0.30^{+0.15}_{-0.11}$    \\     
RCW~79B\tnote{b}  & $15.4^{+1.2}_{-1.2}$ & $4.3^{+0.5}_{-0.6}$  & $15.9^{+0.1}_{-0.1}$ & 2.1$^{+1.2}_{-0.9}$ & $25.8^{+14.3}_{-9.9}$ & $0.52^{+0.29}_{-0.20}$    \\ 
RCW~120A & $21.5^{+1.3}_{-1.3}$  & $3.5^{+0.1}_{-0.1}$  & $16.5^{+0.1}_{-0.1}$ & 4.6$^{+2.0}_{-1.6}$ & $14.4^{+5.2}_{-3.9}$ & $0.29^{+0.10}_{-0.08}$   \\ 
RCW~120B & $16.6^{+1.1}_{-1.1}$  & $3.8^{+0.1}_{-0.1}$ & $16.4^{+0.1}_{-0.1}$ & 5.7$^{+2.8}_{-2.2}$ & $22.5^{+10.0}_{-7.3}$ & $0.45^{+0.20}_{-0.15}$  \\ 
\hline
         & \multicolumn{6}{c}{PDR}      \\
         \cmidrule(lr){2-7}
         & \multicolumn{2}{c}{plane-parallel(wk2020)} & \multicolumn{2}{c}{non-clumpy(kt2020@1~$M_{\odot}$)} & \multicolumn{2}{c}{clumpy(kt2020@1~$M_\odot$)} \\
   \cmidrule(lr){2-3} \cmidrule(lr){4-5} \cmidrule(lr){6-7} 
Source       & $n_{\rm gas}$  & $G_{0}$   & $n_{\rm gas}$  & $G_{0}$ & $n_{\rm gas}$  & $G_{0}$ \\
        & (cm$^{-3}$)    & (Habing)         & (cm$^{-3}$)    & (Habing)         & (cm$^{-3}$) 
        & (Habing) \\
\hline
RCW~79A  & $1.1^{+0.5}_{-0.3}\times 10^{5}$ & 751$^{+268}_{-200}$ &  $3.1^{+2.0}_{-1.1}\times 10^{3}$  & 102$^{+34}_{-30}$  & $1.8^{+1.1}_{-0.7}\times 10^{4}$  & 172$^{+64}_{-48}$ \\     
RCW~79B\tnote{b}  & $7.6^{+4.9}_{-2.6}\times 10^{4}$ & 1957$^{+982}_{-669}$ & $8.3^{+6.6}_{-3.1}\times 10^{3}$ & 86$^{+53}_{-34}$  & $6.4^{+4.0}_{-2.4}\times 10^{3}$  & 253$^{+151}_{-100}$  \\ 
RCW~120A & $2.0^{+0.2}_{-0.2}\times 10^{4}$ & 69$^{+13}_{-11}$ & $4.9^{+2.3}_{-1.8}\times 10^{3}$  & 17$^{+7}_{-5}$  & $2.7^{+1.5}_{-0.9}\times 10^{3}$  & 29$^{+11}_{-9}$ \\ 
RCW~120B & $2.4^{+0.2}_{-0.2}\times 10^{4}$ & 74$^{+14}_{-12}$ & $6.7^{+3.3}_{-2.2}\times 10^{3}$  & 18$^{+7}_{-5}$ & $3.7^{+2.1}_{-1.4}\times 10^{3}$   & 31$^{+10}_{-9}$ \\    
\hline        
\end{tabular}
\begin{tablenotes}
 
        \item[a] The $N_{\rm C^{0}}/N_{\rm CO}$ ratio was derived by scaling the measured $N_{\rm C^{0}}/N_{\rm ^{13}CO}$ ratio by the adopted $^{12}$C/$^{13}$C isotopic ratio of 50.
        \item[b] The large uncertainties of RCW~79B are mainly caused by the low signal-to-noise ratios of [CI] data. The H$_{2}$ number density is not well constrained in the RADEX model.
        \normalsize
\end{tablenotes}
\end{threeparttable}
\end{table*}




\end{document}